\documentclass[superscriptaddress,preprintnumbers,nofootinbib,11pt]{revtex4}

\usepackage[utf8]{inputenc}
\usepackage{graphicx}
\usepackage{color}
\usepackage{multirow}
\usepackage{amsmath,amssymb,amsfonts}
\usepackage{hyperref}
\usepackage{bbm}
\usepackage{url}
\usepackage{slashed}
\usepackage{subfigure}
\usepackage[usenames,dvipsnames]{xcolor}
\usepackage{float} 
\usepackage{epsfig}
\usepackage{graphicx}
\usepackage{euscript}
\usepackage{color}
\usepackage{slashed}
\usepackage{epstopdf}
\usepackage[utf8]{inputenc}
\usepackage[normalem]{ulem}

\usepackage{hyperref}

\newcommand{\be}{\begin{equation}}
\newcommand{\ee}{\end{equation}}
\newcommand{\bea}{\begin{eqnarray}}
\newcommand{\eea}{\end{eqnarray}}

\newcommand{\nn}{\nonumber}

\def \pt{\partial}

\def \tr{\textmd{tr}}
\def \STr{\textmd{STr}}
\def \TT{\textmd{TT}} 
\def \T{\textmd{T}} 
\def \L{\textmd{L}} 
\def \P{\mathcal{P}} 

\def \GN{G_{\textmd{N}}}

\begin{document}

\title{Unimodular quantum gravity: Steps beyond perturbation theory}

\author{Gustavo P. de Brito}
\email{gpbrito@cbpf.br}
\affiliation{Centro Brasileiro de Pesquisas F\'{i}sicas (CBPF),\\ Rua Dr Xavier Sigaud 150, Urca, Rio de Janeiro, RJ, Brazil, CEP 22290-180}
  
\author{Antonio D.~Pereira}
\email{adpjunior@id.uff.br}
\affiliation{Instituto de F\'isica, Universidade Federal Fluminense, Campus da Praia Vermelha, Av. Litor\^anea s/n, 24210-346, Niter\'oi, RJ, Brazil}

\begin{abstract}
The renormalization group flow of unimodular quantum gravity is computed by taking into account the graviton and Faddeev-Popov ghosts anomalous dimensions. In this setting, a ultraviolet attractive fixed point is found. Symmetry-breaking terms induced by the coarse-graining procedure are introduced and their impact on the flow is analyzed. A discussion on the equivalence of unimodular quantum gravity and standard full diffeomorphism invariant theories is provided beyond perturbation theory.
\end{abstract}

\maketitle
\section{Introduction}

The recent direct detection of gravitational waves \cite{Abbott:2016blz} as well as the visualization of a black hole shadow \cite{Akiyama:2019cqa} have enabled a further confirmation of Einstein's field equations as a very accurate description of classical spacetime dynamics even at very strong curvature regimes. Nevertheless, the presence of singularities is ubiquitous in general relativity (GR) and at very short distance scales, quantum fluctuations of spacetime itself must be taken into account. Thus, such a consistent theory of quantum gravity should remove those singularities and have GR as a suitable classical limit. However, the standard quantization of GR with the perturbative continuum quantum field theories techniques leads to a theory which is not perturbatively renormalizable \cite{tHooft:1974toh,Christensen:1979iy,Goroff:1985th}. Consequently, at every loop order in perturbation theory, new counterterms accompanied with free parameters have to be introduced spoiling the predictive power of the underlying quantum field theory. 

Yet, such a theory can be regarded as an effective field theory \cite{Donoghue:1995cz,Burgess:2003jk}. For scales much smaller than the ultraviolet (UV) cutoff, introduced to regularize the theory, infinitely many terms of the effective description are suppressed by inverse powers of the UV cutoff giving rise to a predictive theory. As soon as the UV cutoff scale is reached, the effective field theory breaks down since the suppression mechanism aforementioned does not work anymore. Being a perfectly valid description of the quantum aspects of the gravitational field up to the UV cutoff, the effective field theory perspective shows that there is no \textit{a priori} riddle between the quantum-field theoretic toolbox and GR. As a celebrated result of this fact, quantum gravitational corrections to the Newtonian potential can be computed see, e.g., \cite{Donoghue:1993eb,Donoghue:1994dn}. Nevertheless, at the UV cutoff scale, the effective field theory has to be replaced by a fundamental description of the quantum dynamics of spacetime.

In a path integral quantization, the aimed fundamental description can be pictorially represented by the functional integral\footnote{For future covenience we consider the Euclidean version of the path integral.}
\begin{equation}
\mathcal{Z}_{\mathrm{QGR}} = \int \mathcal{D}g_{\mu\nu}\,\mathrm{e}^{-S_{\mathrm{EH}}}\,,
\label{intro1}
\end{equation}
with
\begin{equation}
S_{\mathrm{EH}}=\frac{1}{16\pi \GN}\int\mathrm{d}^4x\sqrt{g}\left(2\Lambda - R(g)\right)\,,
\label{intro2}
\end{equation}
being the Einstein-Hilbert action with $R(g)$ denoting the Ricci scalar associated with the metric $g_{\mu\nu}$ and $\Lambda$ and $\GN$, the cosmological constant and Newton constant, respectively. The measure $\mathcal{D}g_{\mu\nu}$ as well as the action $S_{\mathrm{EH}}$ are invariant under diffeomorphisms. The choice of the weight $S_{\mathrm{EH}}$ in the partition function is due to its well-grounded description of classical gravitational phenomena by the Einstein-Hilbert action. Such a choice carries over the symmetry group to the measure (in the absence of anomalies, of course). 

However, an equivalent description of the gravitational field dynamics, at the classical level, is provided by unimodular gravity \cite{Anderson:1971pn,vanderBij:1981ym,Buchmuller:1988yn,Buchmuller:1988wx,Unruh:1988in,Henneaux:1989zc,Ellis:2010uc}. In such a theory, the determinant of the metric is a fixed density, i.e., $g \equiv \mathrm{det}\,g_{\mu\nu} = \omega^2$, with $\omega=\omega(x)$ a fixed density. In general, diffeomorphisms will not preserve the determinant of the metric, so unimodular gravity features a reduced symmetry group, since the variation of the metric determinant with respect to a coordinate transformation is constrained by $\delta g = 0\,\Rightarrow\,g^{\mu\nu}\delta g_{\mu\nu}=0$. This requirement imposes a tracelessness condition on the variation of the metric, i.e., $\delta g = 0\,\Rightarrow\,g^{\mu\nu}\delta g_{\mu\nu}=0$. For a coordinate transformation, $\delta g_{\mu\nu} = g_{\nu\alpha}\nabla_\mu \epsilon^\alpha + g_{\mu\alpha}\nabla_\nu \epsilon^\alpha$ for an infinitesimal $\epsilon^\alpha$, this condition translates to $\nabla_{\mu}\epsilon^\mu = 0$. For this reason, such a restricted set of coordinate transformations is called transverse diffeomorphisms\footnote{They are also called special diffeomorphisms, \cite{Ardon:2017atk,Percacci:2017fsy}} \cite{Alvarez:2006uu,Alvarez:2008zw} .

Starting from action \eqref{intro2}, it is possible derive the equations of motion for unimodular gravity by taking variations of the metric subject to the traceless condition. The cosmological constant term does not contribute to the equations of motion since the determinant of the metric is non-dynamical. Besides that, Einstein's field equations can be recovered by means of the Bianchi identities (and imposition of conservation of the energy-momentum tensor when matter is present\footnote{The derived energy-momentum tensor from the unimodular equations of motion is not identically conserved, but can be improved to the same as in GR which shares the same trace-free part with the unimodular's one.}). A term akin to the cosmological constant emerges as an integration constant. Thence, GR and unimodular gravity feature the same classical dynamical content. Conceptually, the role of the cosmological constant is different however. In GR, it appears directly in the Einstein-Hilbert action and couples to the dynamical metric determinant. In unimodular gravity, it is just a constant that shows up in the equations of motion. For this reason, some authors have advocated that unimodular gravity might provide a solution to (one of) the cosmological constant problem(s) (or, at least, a different viewpoint of it), see \cite{vanderBij:1981ym,Buchmuller:1988wx,Wilczek:1983as,Zee:1983jg,Weinberg:1988cp,Henneaux:1989zc,Ellis:2010uc,Alvarez:2005iy,Alvarez:2007nn}, but see also \cite{Padilla:2014yea,Padilla:2015aaa}. 

The path integral of unimodular quantum gravity (UQG) can be represented as 
\begin{equation}
\mathcal{Z}_{\mathrm{UQG}} = \int \mathcal{D}\tilde{g}_{\mu\nu}\,\mathrm{e}^{-S_{\mathrm{UG}}}\,,
\label{intro3}
\end{equation}
where the weight now corresponds to the classical unimodular action
\begin{equation}
S_{\mathrm{UG}} = -\frac{1}{16\pi \GN}\int\mathrm{d}^4x~\omega\, R(\tilde{g})\,,
\label{intro4}
\end{equation}
and the measure is restricted to unimodular metrics $\tilde{g}_{\mu\nu}$. The cosmological constant is absent in the Boltzmann factor of the partition function since the determinant of the metric is non-dynamical, i.e., ``it does not gravitate''. Furthermore, the action and the measure are invariant under \textit{TDiff}.

Albeit classically equivalent, GR and unimodular gravity can differ in the quantum realm. In fact, just from the pictorial representation of their respective path integrals \eqref{intro1} and \eqref{intro3}, it is possible to see that the configuration space for each case is different. However, this question seems to be very subtle, see, e.g., \cite{Ardon:2017atk,Percacci:2017fsy,Padilla:2014yea,Smolin:2009ti,Smolin:2010iq,Alvarez:2015sba,Bufalo:2015wda,Upadhyay:2015fna,Eichhorn:2013xr,Eichhorn:2015bna} for some recent discussions. In both cases, the resulting  quantum field theory is not perturbatively renormalizable and can be treated as an effective field theory below the Planck scale. 

One possible consistent way of providing a UV completion for such quantum field theories is the existence of a UV attractive non-Gaussian fixed point in the renormalization group flow. Such a program, which aims at providing firm ground for this conjecture is the so-called Asymptotic Safety scenario for quantum gravity \cite{Hawking:1979ig,Reuter:1996cp}, see e.g. \cite{Percacci:2017fkn,Eichhorn:2018yfc,Reuter:2019byg,Bonanno:2020bil} for recent reviews. In the case of full $\textit{Diff}$-invariant theories, there are quite compelling evidence for the existence of this fixed point see, e.g.,
\cite{Souma:1999at,Reuter:2001ag,Litim:2003vp,Codello:2006in,Machado:2007ea,Codello:2008vh,Eichhorn:2009ah,Benedetti:2009rx,Benedetti:2009gn,Eichhorn:2010tb,Manrique:2010am,Manrique:2011jc,Christiansen:2012rx,Falls:2013bv,Benedetti:2013jk,Ohta:2013uca,Falls:2014tra,Christiansen:2014raa,Falls:2014zba,Falls:2015qga,Christiansen:2015rva,Ohta:2015efa,Ohta:2015fcu,Knorr:2017mhu,Christiansen:2017bsy,Gies:2015tca,Gies:2016con,Biemans:2016rvp,Denz:2016qks,Falls:2016msz,Falls:2017lst,Falls:2018ylp,deBrito:2018jxt,Knorr:2019atm,Burger:2019upn,Falls:2020qhj}. Meanwhile, hints of the realization of a scale-invariant regime in the UV in the context of unimodular gravity are more scarce, but already found in \cite{Eichhorn:2013xr,Saltas:2014cta,Eichhorn:2015bna,Benedetti:2015zsw}. Therefore, in both cases, a fixed point featuring a finite number of relevant directions\footnote{Such a number correspond to how many free parameters must be fixed in the theory by experimental data. Having a finite number of relevant direction is thus crucial for predictivity.} is found within certain approximations. Conceptually, the space of couplings associated to operators compatible with the symmetries of the corresponding quantum field theories, the theory space, is different for GR and unimodular gravity. A particular example is the cosmological constant. In the theory space spanned by \textit{Diff}-invariant operators, the ``volume operator" which has cosmological constant as its coupling corresponds to a direction in theory space while in the theory space of  \textit{TDiff} theories, it is absent. Consequently, such theories might display a UV attractive fixed point and might fall in different universality classes being thus inequivalent at the quantum level. In \cite{Eichhorn:2013xr,Eichhorn:2015bna,Benedetti:2015zsw}, the fixed points discovered in unimodular quantum gravity within certain truncations feature one relevant direction less with respect to the standard full \textit{Diff}-invariant case very likely associated to the absence of the cosmological constant as a coupling. Nevertheless, the cosmological constant in an intrinsic free parameter of unimodular gravity which arises as an integration constant and, therefore, has to be fixed by initial conditions.

The main (semi-)analytical tool for the quest of non-trivial fixed points in theories of quantum gravity is the functional renormalization group (FRG) \cite{Wetterich:1992yh}. It allows for the computation of the running of the coupling constants beyond the standard perturbative scheme, see \cite{Pawlowski:2005xe,Gies:2006wv,Rosten:2010vm,Dupuis:2020fhh} for some reviews on the FRG. Yet, approximations (truncations) are required for a concrete computation. In this work, tapping on the previous results on asymptotically safe unimodular gravity we improve the previous approximations by following a strategy outlined in \cite{Christiansen:2012rx,Codello:2013fpa}. In particular, we evaluate the anomalous dimensions of the graviton as well as Faddeev-Popov ghosts independently with a view towards the closure of the FRG equation beyond the already employed ``background approximation''. Symmetry-breaking terms of a certain class, generated by the coarse-graining regulator function, are also taken into account. Furthermore, we provide a comparison with full \textit{Diff}-invariant theories quantized in the unimodular gauge. In this gauge, the trace mode of the spin-two fluctuation is removed by a gauge condition, not to be confused with unimodular gravity, where this mode is absent from the beginning and therefore, no compensating Faddeev-Popov ghosts are introduced for this. We find indications that unimodular gauge and gravity are equivalent in a particular sense.

The paper is organized as follows: In Sect.~\ref{sect:setup} we provide a review of technical aspects of the FRG applied to unimodular gravity. Sect.~\ref{Flow2Pt} is devoted to the study of the flow of two-point functions in unimodular gravity and the anomalous dimensions associated to the graviton and Faddeev-Popov ghosts are computed in different schemes. The renormalization group flow and fixed point structure are discussed in Sect.~\ref{RGFlowAndFP} followed by a discussion on the (in)equivalence of unimodular gravity and gauge in Sect.~\ref{Equiv_UGrav-UGauge}. Finally, we collect our conclusions and perspectives. Lenghty expressions and further technical details are left to the appendices. 

\section{Setup}\label{sect:setup}
\noindent

After the seminal work \cite{Reuter:1996cp}, the asymptotic safety program for quantum gravity has found a systematic tool for the search of the non-Gaussian fixed point by means of (semi-)analytical methods\footnote{Alternatively, lattice methods can be employed as in the (Causal) Dynamical Triangulation approach \cite{Loll:2019rdj}.}. The FRG has enabled a substantial progress in the field. Its basic idea lies on a smooth implementation of the Wilsonian renormalization philosophy where the path integral is not performed at once but field modes are integrated out shell by shell. Its central idea is the introduction of a regulator term in the Boltzmann factor of the path integral that suppresses modes with momenta lower than a fixed cutoff $k$. The regulator is quadratic in the quantum fluctuations and has the general structure $\Delta S_k = \int_x \phi(x) \mathbf{R}_k(\Delta) \phi(x)$, where $\phi (x)$ stands for a generic field of the theory. The function $\mathbf{R}_k(\Delta)$ is the kernel responsible to decouple the ``slow modes" that are labelled by momentum scales smaller than $k$. Therefore the bare action $S$ is replaced as  $S[\phi] \mapsto S_k[\phi] = S[\phi] + \Delta S_k[\phi]$ and a scale-dependent generating functional is defined by
\begin{align}
Z_k[J] = \int \mathcal{D}\phi \, \mathrm{e}^{-S_k[\phi]+\int_x J\cdot \phi}\,,
\end{align}
where $J$ stands for an external source. The main object in the FRG is the flowing effective action $\Gamma_k$ defined as a modified Legendre transform of the scale-dependent generating functional of connected Feynman diagrams $W_k[J]$ ($ = \ln Z_k[J]$), namely,
\begin{align}
\Gamma_k[\varphi] = \sup_J \bigg( \int\mathrm{d}^4x\, J \cdot \varphi - W_k[J] - \Delta S_k[\varphi] \bigg)\,,
\end{align}
where $\varphi \equiv \langle \phi\rangle_J$. The flowing effective action $\Gamma_k$ interpolates between the microscopic (bare) action $\Gamma_{k\to\Lambda} = S_\Lambda$ (with $\Lambda$ being a UV cutoff) and the full effective action (i.e., the generating functional of the one-particle-irreducible Feynman diagrams) $\Gamma_{k\to0} = \Gamma$. Remarkably, the effective average action $\Gamma_k$ satisfies an exact flow equation which has an one-loop structure\footnote{Thanks to the quadratic dependence of the regulator function on the fields.}, the Wetterich equation \cite{Wetterich:1992yh}, formally written as
\begin{align} \label{Flow_Eq}
\pt_t \Gamma_k = \frac{1}{2} \STr \left[ \pt_t \textbf{R}_k \,(\Gamma^{(2)}_k + \textbf{R}_k)^{-1} \right] ,
\end{align}
where $\pt_t \equiv k \pt_k$, $\Gamma^{(2)}_k \equiv \delta^2 \Gamma_k/\delta \varphi \delta \varphi$ and $\STr$ denotes the supertrace which takes into account appropriate numerical factors depending on the nature of the fields.
Although \eqref{Flow_Eq} is formally an exact equation, practical applications of the FRG require approximations. An useful strategy is to consider a truncation for $\Gamma_k$ where a particular set of operators is taken into account\footnote{The effective average action should contain all operators compatible with the symmetries of the deformed action $S_k$.}. When applied to quantum gravity, the FRG is combined with the background field method \cite{Abbott:1981ke}. The background metric $\bar{g}_{\mu\nu}$ defines a ``momentum scale" through the eigenvalues of the background Laplacian and field modes are organized accordingly. On top of that, gauge symmetries are fixed by the standard quantum-field theoretic tools leading to the insertion of gauge-fixing terms as well as Faddeev-Popov ghosts. For reviews of the FRG applied to gauge theories and gravity, we refer to, e.g., \cite{Pawlowski:2005xe,Gies:2006wv,Dupuis:2020fhh}. In this work, we focus on unimodular gravity and, in the following, we highlight some peculiarities of such a theory and how the FRG should be adapted in this case.

Unimodular gravity is characterized by a fixed non-dynamical metric determinant, i.e.,
\begin{align}\label{unimodularity}
\det g_{\mu\nu} = \omega^2 \,,
\end{align}
where $\omega$ denotes a fixed scalar density. While general relativity is constructed upon the assumption of covariance under \textit{Diff} transformations, the unimodular condition \eqref{unimodularity} selects only the subgroup of volume preserving \textit{Diff} transformations, \textit{TDiff}, since the volume preserving transformation acting on the metric is given by
\begin{align}
\delta_{\epsilon_\T} g_{\mu\nu}(x) 
= \mathcal{L}_{\epsilon_\T} g_{\mu\nu}(x) 
= g_{\mu\alpha}\nabla_{\nu} \epsilon_\T^{\alpha}(x) + g_{\nu\alpha}\nabla_{\mu} \epsilon_\T^{\alpha}(x) ,
\end{align}
with transverse generators $\epsilon_\T^\mu$, i.e., $\nabla_\mu \epsilon^\mu_\T=0$.

The functional quantization of unimodular gravity is based on the formal definition of a path integral over metric configurations satisfying \eqref{unimodularity}. There are several different ways of imposing the constraint \eqref{unimodularity} at the level of the path integral\footnote{It is not clear if the resulting quantum theories are equivalent.}, see, e.g., \cite{Eichhorn:2013xr,Ardon:2017atk,Percacci:2017fsy,Smolin:2009ti,Alvarez:2015sba,Bufalo:2015wda,Upadhyay:2015fna,Saltas:2014cta,Baulieu:2020obv}. In this paper we follow the same strategy as in \cite{Eichhorn:2013xr} which combines the background field method with an exponential split for the metric. The exponential parameterization, first introduced in the context of $2+\epsilon$ expansion \cite{Kawai:1992np,Kawai:1993fq,Kawai:1993mb} and later applied in the FRG framework \cite{Eichhorn:2013xr,Eichhorn:2015bna,Nink:2014yya,Gies:2015tca,Percacci:2015wwa,Labus:2015ska,Ohta:2015efa,Dona:2015tnf,Benedetti:2015zsw,Ohta:2015fcu,Ohta:2016npm,Ohta:2016jvw,deBrito:2018jxt,Alkofer:2018fxj,Alkofer:2018baq,deBrito:2019umw}, takes the form
\begin{align}
g_{\mu\nu} = \bar{g}_{\mu\alpha} [\exp (\kappa\, h^{\cdot}_{\,\,\,\,\cdot})]^{\alpha}_{\,\,\,\,\nu} 
= \bar{g}_{\mu\nu} + \kappa \, h_{\mu\nu} + \sum_{n=2}^\infty \frac{\kappa^n}{n!} h_{\mu\alpha_1} 
\cdots h^{\alpha_{n-1}}_{\,\,\,\nu} \,,
\end{align}
where $\bar{g}_{\mu\nu}$ represents a background metric, which is chosen to be maximally symmetric in this work - and unimodular - $h_{\mu\nu}$ denotes the fluctuation field and we have defined $\kappa = (32\pi \GN)^{1/2}$, where $\GN$ denotes the dimensionful Newton constant. The main advantage of using the exponential parameterization is the fact that one can express the metric determinant as $\det g_{\mu\nu} = \det \bar{g}_{\mu\nu} \, \mathrm{e}^{h^\tr}$ (with $h^\tr=\bar{g}^{\mu\nu} h_{\mu\nu}$) and, therefore, the unimodularity condition \eqref{unimodularity} can be easily implemented by combining $\det \bar{g}_{\mu\nu} = \omega^2$ with the tracelessness condition $h^\tr=0$. From this perspective the functional quantization of unimodular gravity translates into a path integral over fluctuations $h_{\mu\nu}$ which are traceless. Additionally, a novel subtle point regarding the path integral for unimodular gravity was pointed out in \cite{Ardon:2017atk,Percacci:2017fsy}. In fact, when applying the Faddeev-Popov quantization in this case, the identification of the volume of the gauge group \textit{TDiff} requires the introduction of an extra determinant factor of a background Laplacian $\mathrm{Det}^{1/2}~\Delta_0$ in the path integral measure. We elaborate a bit more on that in App.~A. Therefore, in order to reproduce correctly the one-loop results and since this is genuinely part of the Faddeev-Popov construction of unimodular gravity, we include such a term directly when ``deriving" the flow equation for unimodular gravity. Consequently, after a proper regularization of such a determinant factor, the standard derivation of the flow equation leads to
\begin{align}\label{flowequnim}
\pt_t \Gamma_k = \frac{1}{2} \STr \left[ \pt_t \textbf{R}_k \,(\Gamma^{(2)}_k + \textbf{R}_k)^{-1} \right] - \frac{1}{2}\mathrm{Tr} \left[ \!\frac{}{}\pt_t R_k \,(\Delta_0 + R_k)^{-1} \right]  \,,
\end{align}
where $R_k$ is the kernel of the regulator introduced for the scalar determinant arising from the measure. Some comments are important: In principle, one could ignore completely the path integral perspective and just look at the flow equation and construct a theory based on a theory space defined by \textit{TDiff}. In this case, such an extra term would not be present and from the point of view of the FRG, this might not be taken as a serious issue since it is just a different choice of truncation. However, if one wants to match one-loop calculations perfomed with standard functional integral techniques and those obtained with the FRG, then the introduction of such a term seems to be crucial. Hence, this seems to be already a good reason for its inclusion. Another important comment is that it is by no means clear if the Faddeev-Popov trick is well-defined at the non-perturbative level due to the possible existence of Gribov copies. Nevertheless, assuming the standard perturbative Faddeev-Popov method, the extra contribution appears to be purely associated to the background. This means that for a background approximation with the FRG, this term will contribute while for the computation of the flow of correlation functions, this term is completely irrelevant since functional derivatives with respect to quantum flucuations when acting on it give zero.

The flowing action $\Gamma_k$ of unimodular gravity includes all possible terms compatible with the deformed \textit{TDiff} symmetry due to the introduction of the regulator function. Typically, the presense of the regulator spoils the invariance of the action under the original gauge group as well as under ``split symmetry" between background and fluctuation fields\footnote{Being a gauge theory, the local symmetry under \textit{TDiff} is broken by the gauge-fixing term and replaced by BRST invariance. Therefore, more precisely, the regulator deforms the BRST symmetry. Moreover, the background field method leads to the split of the full field in a background piece and a fluctuation part, but there is a symmetry - which is known as split symmetry - that restricts the functional form of the effective action in such a way that it depends of the appropriate combination of background and fluctuations fields. Again, the gauge-fixing term treats the background and the flucutations fields in such a way that such a symmetry is broken. Nevertheless, such a breaking comes in the form of a BRST-exact term. For the regulator, however, such a breaking is explicit and not BRST exact.}. Consequently, the associated Ward identities are modified by the introduction of regulator-dependent terms at finite $k$. In this sense, symmetry preservation at $k=0$ requires symmetry breaking terms at $k\neq 0$. So far, most of the works done with the FRG in quantum gravity considered truncations respecting the original symmetry at $k\neq 0$. However, in the last few years there was important progress regarding the inclusion of symmetry breaking terms in truncations \cite{Dona:2015tnf,Eichhorn:2018akn,Eichhorn:2018ydy,Eichhorn:2018nda}.

The main purpose of this paper is to investigate the renormalization group flow of the graviton and Faddeev-Popov ghosts 2-point functions in unimodular gravity. In this sense, we employ the strategy put forward in \cite{Christiansen:2012rx,Codello:2013fpa}  which is based on the vertex expansion approach for the FRG. The basic idea is to expand $\Gamma_k$ in terms of its proper vertices. Schematically, the vertex expansion takes the form\footnote{Each functional derivative is associated to a space-time variable and the integral represents a collective integration over all such variables.}
\begin{align}
\Gamma_{k}[\varphi\,;\bar{g}] = 
\sum_{n} \frac{1}{n!} 
\int \Gamma_{k,\,A_1\,\cdots A_n}^{(n)}[\bar{g}] \,\varphi^{A_n}\,\cdots\,\varphi^{A_1}\, ,
\end{align}
with vertices defined according to
\begin{align}
\Gamma_{k,\,A_1\,\cdots A_n}^{(n)}[\bar{g}] = 
\frac{\delta^{n} \Gamma_k}{ \delta\varphi^{A_1}\,\cdots\delta\varphi^{A_n} }\bigg|_{\varphi = 0} \,,
\end{align}
where we have used the ``super-field'' notation $\varphi_{A} = (h_{\mu\nu}, \bar{c}_\mu \, ,c^\mu\,,b_\mu)$. Note that, besides the fluctuation field $h$, the flowing action $\Gamma_k$ also has a functional dependence on the Faddeev-Popov ghosts $c^\mu$ and $\bar{c}_\mu$, as well as on the Lautrup-Nakanishi field $b_{\mu}$. Typically, the Lautrup-Nakanishi field is not included as part of the configuration space in FRG truncations in quantum gravity. The advantage for the inclusion of such a field in the present paper will be discussed in Sect. \ref{Equiv_UGrav-UGauge}.

In order to define the truncated vertices we follow the same construction employed in \cite{Dona:2015tnf,Eichhorn:2018nda}. The idea is to define a ``seed'' truncation $\hat{\Gamma}$ which is used to extract the tensorial structure that enters in the vertex expansion of the flow equation. In the present paper we choose $\hat{\Gamma}$ to take the form
\begin{align}\label{seed_truncation}
\hat{\Gamma}[h,c,\bar{c},b;\bar{g}] = 
\hat{\Gamma}_{\textmd{UG}}[g(h\,;\bar{g})] + \hat{\Gamma}_{\textmd{g.f.}}[h,c,\bar{c},b\,;\bar{g}] + 
\hat{\Gamma}_{m^2}[h;\bar{g}] \,.
\end{align}
The first term, $\hat{\Gamma}_{\textmd{UG}}$, includes only contributions which are invariant under the original \textit{TDiff} symmetry. Our ``seed" truncation for $\hat{\Gamma}_{\textmd{UG}}$ corresponds to the unimodular version of the Einstein-Hilbert action in four dimensions,
\begin{align}
\hat{\Gamma}_{\textmd{UG}}[g(h\,;\bar{g})] = -\frac{1}{16\pi \GN} \int_x \omega\, R(g(h\,;\bar{g})) \,, 
\end{align}
where we have defined the compact notation $\int_x = \int \mathrm{d}^4x $. The second term in \eqref{seed_truncation} corresponds to the gauge-fixing action obtained through the Faddeev-Popov procedure\footnote{We refer to App.~\ref{FlowEq_UG} for comments on the Faddeev-Popov procedure in the unimodular setting.}, which replaces gauge invariance by BRST symmetry, namely
\begin{align}
\hat{\Gamma}_{\textmd{g.f.}}[h,c,\bar{c},b\,;\bar{g}] = 
\int_x  \omega\, \bar{g}^{\mu\nu} \,b_{\mu} F_{\nu}^\T[h\,;\bar{g}] -
\frac{\alpha\,}{2} \int_x \omega\,\bar{g}^{\mu\nu} \,b_{\mu} b_{\nu} + 
\int_x \omega\,\bar{c}_{\mu} \,{{\mathcal{M}}^{\mu}}_{\nu}[h\,;\bar{g}] \,c^{\nu}  \,,
\end{align}
where $\alpha$ represents a gauge parameter. We use the transverse gauge condition given by $F_{\mu}^\T[h\,;\bar{g}] = \sqrt{2}\,\P_{\T,\mu}^{\quad\,\nu} \,\bar{\nabla}^\alpha h_{\nu\alpha}$, where $\P_{\T,\mu}^{\quad\,\nu} = \delta_\mu^\nu - \bar{\nabla}_\mu (\bar{\nabla}^2)^{-1} \bar{\nabla}^\nu$ is the transverse projector. The Faddeev-Popov operator is defined according to usual relation,
\begin{align}
{{\mathcal{M}}^{\mu}}_{\nu}[h\,;\bar{g}] \,c^{\nu} = 
\bar{g}^{\mu\nu}\frac{\delta F_\nu^\T[h\,;\bar{g}]}{\delta h_{\alpha\beta}} \delta_c^Q h_{\alpha\beta} \,.
\end{align}
In the context of unimodular gravity, the Faddeev-Popov ghost %and the Lautrup-Nakanishi field 
is constrained by a transversality condition, namely $\nabla_\mu c^\mu = 0$ %= \nabla^\mu \bar{c}_\mu = 0$ and $\nabla^\mu b_\mu = 0$, 
, which follows from the transverse nature of $\epsilon^\mu_\T$. Thanks to the unimodularity condition, we can recast the transversality constraint in terms for the background covariant derivative\footnote{It has to be understood that the replacement of covariant derivative by background covariant derivatives means $\nabla_\mu c^\mu = \bar{\nabla}_\mu\left(g^{\mu\nu}c_\nu\right)$, i.e., the ghost field as the fundamental variable in our calculation must be a contravariant vector.}, i.e., $\bar{\nabla}_\mu c^\mu =0$.  %= \bar{\nabla}^\mu \bar{c}_\mu =0$ and $\bar{\nabla}^\mu b_\mu =0$\ToCite. 
The nonlinear nature of the exponential decomposition for the full metric requires some attention regarding the gauge fixing sector since its nonlinear character induces infinitely many terms in the \textit{TDiff} transformation applied to the fluctuation field,
\begin{align}
\delta_c^Q h_{\mu\nu} = \sum_{n=0}^\infty Y_{\mu\nu}^{(n)} \,,
\end{align}
where $Y_{\mu\nu}^{(n)}$ denotes contributions of order $\mathcal{O}(h^n)$. These contributions can be computed by means of the following recursive relations (see App. B in Ref. \cite{Eichhorn:2017sok})
\begin{subequations}
	\begin{align}
	Y_{\mu\nu}^{(0)}(x)= \mathcal{L}_c \bar{g}_{\mu\nu}(x) \,,
	\end{align}
	and
	\begin{align}
	Y^{(n)}_{\mu\nu}(x)= \mathcal{L}_c X_{\mu\nu}^{(n)}(x) - 
	\sum_{r=0}^{n-1} \int_y \frac{\delta X_{\mu\nu}^{(n-r+1)}(x)}{\delta h_{\alpha\beta}(y)} Y_{\alpha\beta}(y) \, ,  
	\end{align}
\end{subequations}
for $n\geq 1$, where $X_{\mu\nu}^{(n)} = \frac{1}{n!} h_{\mu\alpha_1} \cdots h^{\alpha_{n-1}}_{\,\,\,\nu}$. As a consequence, the ghost sector in unimodular gravity exhibit higher-order ghost-graviton vertices that are not present in FRG truncations based on the linear split of the metric\footnote{Actually, this is not a property of unimodular gravity, but rather of the exponential parameterization.}, see, e.g., \cite{Eichhorn:2010tb}.

The last term in \eqref{seed_truncation} extends the truncated theory space to the sector of symmetry-breaking operators induced by the regulator. In this paper, as a first step in unimodular gravity, we consider a simple mass-like term for the fluctuation field,
\begin{align}
\hat{\Gamma}_{m_h^2}[h;\bar{g}]  = \frac{m_h^2}{2} \int_x \bar{g}^{\mu\alpha}\bar{g}^{\nu\beta} h_{\mu\nu} h_{\alpha\beta} \,,
\end{align}
where $m_h$ denotes a mass parameter\footnote{The introduction of such a term should not be confused with the inclusion of a massive graviton. The regulator deforms the Slavnov-Taylor identities and a mass-term like this cannot be avoided. However, at vanishing $k$, the standard BRST symmetry is recovered and such terms should vanish.}. As we are going to see later, even if not present in the original truncation, this term is generated by the flow of the 2-point function $\delta^2\Gamma_k/\delta h^2$. 

To extract the truncated vertices $\Gamma_{k,\,A_1\,\cdots A_n}^{(n)}$ from the ``seed'' truncation, we expand $\hat{\Gamma}_{\textmd{UG}}$ and $\hat{\Gamma}_{\textmd{g.f.}}$ up to order $\mathcal{O}(h^4)$ and $\mathcal{O}(h^2)$, respectively. For practical calculations we set the background metric to be flat $\bar{g}_{\mu\nu} = \delta_{\mu\nu}$. In this case, it is convenient to work in Fourier space. We note that higher-order terms in the fluctuation field do not give any contribution to the results presented in this paper. The truncated vertices $\Gamma_{k,\,A_1\,\cdots A_n}^{(n)}$ are, then, obtained by dressing the ``seed'' vertices $\hat{\Gamma}_{A_1\,\cdots A_n}^{(n)}$ with tensor structures $[\mathcal{Z}_{k,\varphi}(p)^{1/2}]^{B}_{\,\,\,\, A}$. In such a case, we define 
\begin{align}\label{dressed_vertices}
\Gamma_{k,\,A_1\,\cdots A_n}^{(n)}(\textbf{p}) = 
[\mathcal{Z}_{k,\varphi_1}(p_1)^{1/2}]^{B_1}_{\quad A_1}\cdots \,[\mathcal{Z}_{k,\varphi_n}(p_n)^{1/2}]^{B_n}_{\quad A_n}\,
\hat{\Gamma}_{B_1\,\cdots B_n}^{(n)}(\textbf{p}) \Big|_{\GN \,\mapsto\, G_{k,\textmd{N}}}\,,
\end{align}
with $\textbf{p}=(p_1,\,\cdots,p_{n-1})$ (note that $p_n= -( p_1 + \cdots+ p_{n-1})$ due to momentum conservation) and $G_{k,\textmd{N}}$ denotes the scale-dependent Newton's coupling. The relevant tensor structures are defined according to
\begin{subequations}
	\begin{align}
	[\mathcal{Z}_{k,h}(p)^{1/2}]^{\mu\nu}_{\quad \alpha\beta} =
	\,\,Z_{k,\TT}^{1/2} \, [\mathcal{P}_{\TT}(p)]^{\mu\nu}_{\quad \alpha\beta}
	+ \,\,Z_{k,\xi}^{1/2} \, [\mathcal{P}_{\xi}(p)]^{\mu\nu}_{\quad \alpha\beta} +
	Z_{k,\sigma}^{1/2} \, [\mathcal{P}_{\sigma}(p)]^{\mu\nu}_{\quad \alpha\beta} \,,
	\end{align}
	\begin{align}
	[\mathcal{Z}_{k,\bar{c}}(p)^{1/2}]^{\mu}_{\,\,\,\, \nu}  =
	[\mathcal{Z}_{k,c}(p)^{1/2}]^{\mu}_{\,\,\,\, \nu}  =
	Z_{k,c}^{1/2} \, [\mathcal{P}_{\T}(p)]^{\mu}_{\,\,\,\, \nu} \,,
	\end{align}
	\begin{align}
	[\mathcal{Z}_{k,b}(p)^{1/2}]^{\mu}_{\,\,\, \nu} =
	Z_{k,b}^{1/2} \, [\mathcal{P}_{\T}(p)]^{\mu}_{\,\,\, \nu} \,.
	\end{align}
\end{subequations}
where $Z_{k,\TT}$, $Z_{k,\xi}$, $Z_{k,\sigma}$, $Z_{k,c}$ and $Z_{k,b}$ correspond to the wave-function renormalization factors of the indicated fields. In the graviton sector, we have used the projectors $\mathcal{P}_{\TT}(p)$, $\mathcal{P}_{\xi}(p)$ and $\mathcal{P}_{\sigma}(p)$ defined on the York-basis (see App. \ref{Proj_Flat}). Moreover, $\mathcal{P}_{\T}(p)$ correspond to the transverse projector acting on vector fields. The use of different pre-factors in the expansion of $\mathcal{Z}_{k,h}(p)^{1/2}$ account for possible symmetry breaking effects induced by the FRG regulator. As a further step towards the inclusion of symmetry deformation contributions, we also redefine the mass parameter $m_h^2$, appearing in the graviton 2-point function, according to $m_h^2 \mapsto m_{k,\TT}^2$, $m_h^2 \mapsto m_{k,\xi}^2$ and $m_h^2 \mapsto -\frac{1}{2} m_{k,\sigma}^2$ for the different tensorial sectors defined in terms of the projectors $\mathcal{P}_{\TT}(p)$, $\mathcal{P}_{\xi}(p)$ and $\mathcal{P}_{\sigma}(p)$. In principle, the gauge-fixing parameter $\alpha$ is also allowed to run, therefore, we replace $\alpha\mapsto \alpha_k$.

For the FRG regulator function, in the present paper we consider the following prescription 
\begin{align}
\textbf{R}_{k,A_1 A_2}(p) = 
[\mathcal{Z}_{k,\varphi_1}(p)^{1/2}]^{B_1}_{\quad A_1} \,
[\mathcal{Z}_{k,\varphi_2}(p)^{1/2}]^{B_2}_{\quad A_2} \,
\left( \hat{\Gamma}_{B_1 B_2}^{(2)}(p)\Big|_{p_\mu\mapsto (1+r_k)^{1/2}p_\mu} 
- \hat{\Gamma}_{B_1 B_2}^{(2)}(p)\right) \,,
\end{align}
where $r_k=r_k(p^2)$ denotes the shape function. Here we consider the Litim's regulator $r_k(p^2) = (k^2/p^2-1) \theta(k^2/p^2-1)$, \cite{Litim:2000ci,Litim:2001up}.

\section{Flow of the 2-point function in Unimodular Gravity \label{Flow2Pt}} 

The flow of the 2-point function $\Gamma_k^{(2)}$ can be obtained by acting with two functional derivatives w.r.t. $\varphi$ on the FRG equation. In general, the flow equation for $\Gamma_k^{(2)}$ reads\footnote{The extra ``scalar" trace arising from the measure as described in App.~\ref{FlowEq_UG} drops upon the action of functional derivatives.}
\begin{align}\label{Flow_Eq_2point}
\pt_t \Gamma_{k}^{(2)} = 
-\frac{1}{2} \STr\Big( \textbf{G}_k\, \Gamma_{k}^{(4)} \, \textbf{G}_k \, \pt_t \textbf{R}_k \Big)
+ \STr\Big( \textbf{G}_k \, \Gamma_k^{(3)} \, \textbf{G}_k  \, \Gamma_k^{(3)} \, \textbf{G}_k \, \pt_t \textbf{R}_k \Big) \, ,
\end{align}
where $\textbf{G}_k = (\Gamma_k^{(2)} + \textbf{R}_k)^{-1}|_{\varphi=0}$ denotes the dressed propagator. For the truncation we are considering, the 2-point functions are
\begin{subequations}
	\begin{align}
	\frac{\delta^2\Gamma_k[\varphi]}{\delta h_{\mu\nu}(-p)\delta h_{\alpha\beta}(p)}\bigg|_{\varphi=0} = 
	&\,Z_{k,\TT}\,(p^2+m_{k,\TT}^2)\, \mathcal{P}_\TT^{\mu\nu\alpha\beta}(p)  \nonumber \\
	+&\,Z_{k,\xi}\, m_{k,\xi}^2\, \mathcal{P}_\xi^{\mu\nu\alpha\beta}(p) -
	\frac{1}{2} Z_{k,\sigma}\,(p^2+m_{k,\sigma}^2)\, \mathcal{P}_\sigma^{\mu\nu\alpha\beta}(p) \,,
	\end{align}
	\begin{align}
	\frac{\delta^2\Gamma_k[\varphi]}{\delta h_{\mu\nu}(-p)\delta b_{\alpha}(p)}\bigg|_{\varphi=0} = \frac{1}{2i}
	Z_{k,\xi}^{1/2} Z_{k,b}^{1/2} \, 
	\left( p^\mu \mathcal{P}_\T^{\nu\alpha}(p) + p^\nu \mathcal{P}_\T^{\mu\alpha}(p)  \right) \,,
	\end{align}
	\begin{align}
	\frac{\delta^2\Gamma_k[\varphi]}{\delta b_{\mu}(-p)\delta b_{\alpha}(p)}\bigg|_{\varphi=0} = 
	-\alpha_k \,Z_{k,b}\, \mathcal{P}_\T^{\mu\alpha}(p) \,,
	\end{align}
	\begin{align}
	\frac{\delta^2\Gamma_k[\varphi]}{\delta c_{\mu}(-p)\delta \bar{c}_{\alpha}(p)}\bigg|_{\varphi=0} = 
	-\sqrt{2} \,Z_{k,c}\,p^2\, \mathcal{P}_\T^{\mu\alpha}(p) \,.
	\end{align}
\end{subequations}

Before we proceed with the main results of this paper, let us add a brief remark regarding the running of the gauge-fixing parameter $\alpha_k$. An interesting feature of the inclusion of Lautrup-Nakanishi fields in the FRG truncation is the possibility of extracting the flow of $\alpha_k$ directly from the 2-point function $\delta^2 \Gamma_k/\delta b^2$. The r.h.s. of the flow equation for $\delta^2 \Gamma_k/\delta b^2$ involves 3- and 4-point vertices containing at least one functional derivative w.r.t. the Lautrup-Nakanishi field. However, vertices with these features are not present in the truncation we are considering. In such a case, the r.h.s. of for the flow equation for $\delta^2 \Gamma_k/\delta b^2$ vanishes. As a consequence, the running of $\alpha_k$ can be readily extracted from
$\pt_t\left(\delta^2 \Gamma_k/\delta b^2\right) = 0$, resulting in
\begin{align}
\pt_t\alpha_k = \alpha_k \,\eta_b\,,
\end{align}
where we have defined $\eta_b = - Z_{k,b}^{-1} \pt_t Z_{k,b}$. Since $\pt_t\alpha_k $ is proportional to $\alpha_k$ itself, the Landau gauge choice $\alpha_k = 0$ turns out to be a fixed point. In this sense, we can set $\alpha_k = 0$ along the calculation. The running of $\alpha_k$ has been explored in \cite{Knorr:2017fus}, leading to the same conclusion as in the standard ASQG setting based in the quantization \textit{Diff}-invariant theories. 

At the practical level, the Landau gauge choice simplifies the analysis performed in this paper. In particular, due to the choice $\alpha_k = 0$, the mass parameter $m_{k,\xi}^2$ and the wave-function renormalization factors $Z_{k,\xi}$ and $Z_{k,b}$ do not feedback in the flow of the graviton and ghost 2-point functions. This feature is a consequence of the regularized graviton propagator that takes the form
\begin{align}
\textbf{G}_{k,hh}^{\mu\nu\alpha\beta}(p) = 
\frac{ \mathcal{P}_\TT^{\mu\nu\alpha\beta}(p) }{Z_{k,\TT}\,((1+r_k(p^2))\,p^2+m_{k,\TT}^2)}  -
\frac{ 2\,\mathcal{P}_\sigma^{\mu\nu\alpha\beta}(p) }{Z_{k,\sigma}\,((1+r_k(p^2))\,p^2+m_{k,\sigma}^2)} \,,
\end{align}
in the Landau gauge. For this reason, in the present paper, we focus our attention in the flow of $m_{k,\TT}^2$, $m_{k,\sigma}^2$, $Z_{k,\TT}$, $Z_{k,\sigma}$ and $Z_{k,c}$, with $\alpha_k=0$. The other relevant propagator for our analysis is the ghost propagator,
\begin{align}
\textbf{G}_{k,c\bar{c}}^{\mu\nu}(p) = 
-\frac{1}{\sqrt{2} \,Z_{k,c}\,(1+r_k(p^2)) \,p^2}\,\mathcal{P}_\T^{\mu\nu}(p) \,.
\end{align}
For the sake of completeness, we also include the dressed propagators involving the Lautrup-Nakanishi field
\begin{subequations}
	\begin{align}
	\textbf{G}_{k,hb}^{\mu\nu\alpha}(p) = 
	\frac{1}{\sqrt{2}\,i\,Z_{k,b}^{1/2} Z_{k,\xi}^{1/2} \,(1+r_k(p^2))^{1/2} \,p^2}
	\left( \mathcal{P}_\T^{\mu\alpha}(p) \,p^\nu + \mathcal{P}_\T^{\nu\alpha}(p) p^\mu \right) \,,
	\end{align}
	\begin{align}
	\textbf{G}_{k,bb}^{\mu\nu}(p) = 
	-\frac{m_{k,\xi}^2}{Z_{k,b} \,(1+r_k(p^2)) \,p^2} \,\mathcal{P}_\T^{\mu\nu}(p) \,.
	\end{align}
\end{subequations}

Here we report on results obtained through a derivative expansion. We use the flow equation \eqref{Flow_Eq_2point} to compute the anomalous dimensions $\eta_\TT = -Z_{k,\TT}^{-1}\pt_t Z_{k,\TT}$, $\eta_\sigma = -Z_{k,\sigma}^{-1}\pt_t Z_{k,\sigma}$ and $\eta_c = -Z_{k,c}^{-1}\pt_t Z_{k,c}$ as well as the running of the mass parameters $m_{k,\TT}^2$ and $m_{k,\sigma}^2$. To extract the $\eta$'s and $\pt_t m^2$'s from Eq.\eqref{Flow_Eq_2point} we apply the following projection rules
\begin{subequations}
	\begin{align}\label{Proj_eta_TT}
	\eta_\TT = -\frac{1}{5 \, Z_{k,\TT}}
	\left[\frac{\pt}{\pt p^2} 
	\left(  \mathcal{P}_\TT^{\mu\nu\alpha\beta}(p) 
	\frac{\delta^2\pt_t \Gamma_k[\varphi]}{\delta h^{\mu\nu}(-p) \delta h^{\alpha\beta}(p)}\bigg|_{\varphi=0}
	\right) \right]_{p^2=0} \,,
	\end{align}
	\begin{align}\label{Proj_eta_sigma}
	\eta_\sigma = \frac{2}{Z_{k,\sigma}}
	\left[\frac{\pt}{\pt p^2} 
	\left(  \mathcal{P}_\sigma^{\mu\nu\alpha\beta}(p) 
	\frac{\delta^2\pt_t \Gamma_k[\varphi]}{\delta h^{\mu\nu}(-p) \delta h^{\alpha\beta}(p)}\bigg|_{\varphi=0}
	\right) \right]_{p^2=0} \,,
	\end{align}
	\begin{align}\label{Proj_eta_Gh}
	\eta_c = \frac{1}{3\,\sqrt{2}\,Z_{k,c}}
	\left[\frac{\pt}{\pt p^2} 
	\left(  \mathcal{P}_\T^{\mu\nu}(p) 
	\frac{\delta^2\pt_t \Gamma_k[\varphi]}{\delta c^{\mu}(-p) \delta \bar{c}^{\,\nu}(p)}\bigg|_{\varphi=0}
	\right) \right]_{p^2=0} \,,
	\end{align}
\end{subequations}

\begin{subequations}
	\begin{align}\label{Proj_m_TT}
	\pt_t m^2_{k,\TT} = \eta_\TT\,m^2_{k,\sigma} + \frac{1}{5 \, Z_{k,\TT}}
	\left(  \mathcal{P}_\TT^{\mu\nu\alpha\beta}(p) 
	\frac{\delta^2\pt_t \Gamma_k[\varphi]}{\delta h^{\mu\nu}(-p) \delta h^{\alpha\beta}(p)}\bigg|_{\varphi=0}
	\right)_{p^2=0} \,,
	\end{align}
	\begin{align}\label{Proj_m_sigma}
	\pt_t m^2_{k,\sigma} = \eta_\sigma\,m^2_{k,\sigma} -\frac{2}{Z_{k,\sigma}}
	\left(  \mathcal{P}_\sigma^{\mu\nu\alpha\beta}(p) 
	\frac{\delta^2\pt_t \Gamma_k[\varphi]}{\delta h^{\mu\nu}(-p) \delta h^{\alpha\beta}(p)}\bigg|_{\varphi=0}
	\right)_{p^2=0} \,.
	\end{align}
\end{subequations}
The running for the corresponding dimensionless mass parameters ($\tilde{m}_{k,i}^2 = k^{-2} m_{k,i}^2$) can be obtained by the simple formula $\pt_t \tilde{m}_{k,i}^2 = -2\tilde{m}_{k,i}^2+k^{-2} \pt_t m_{k,i}^2$. The explicit results for $\eta_i$'s and $\pt_t \tilde{m}_{k,i}^2$'s are reported in the App. \ref{Explicit}. 

\begin{figure}[htb!]
	\begin{center}
		\includegraphics[height=1.5cm]{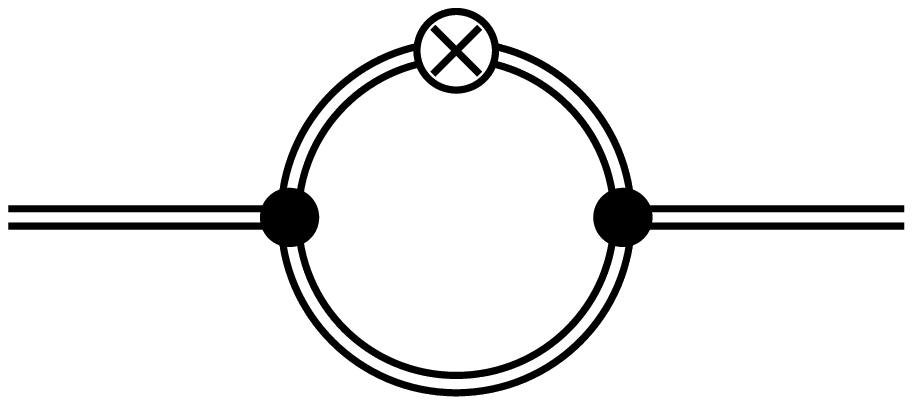} \qquad
		\includegraphics[height=1.5cm]{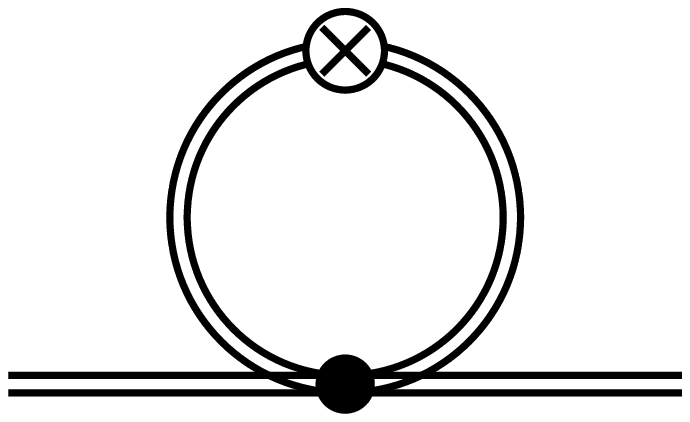} \qquad
		\includegraphics[height=1.5cm]{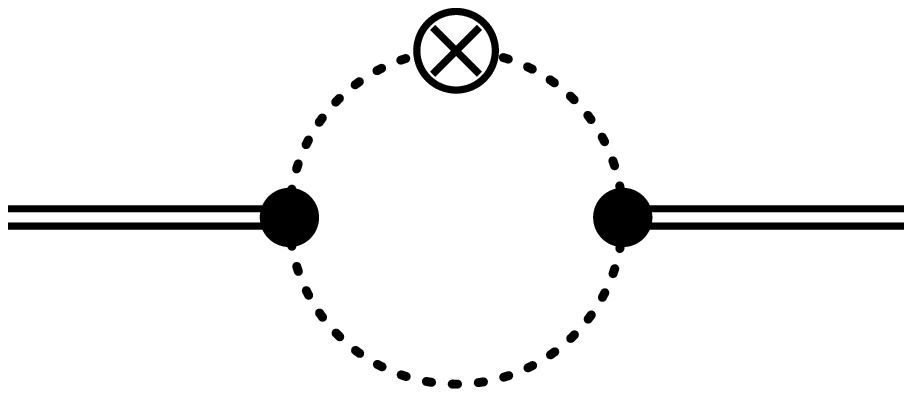} \qquad
		\includegraphics[height=1.5cm]{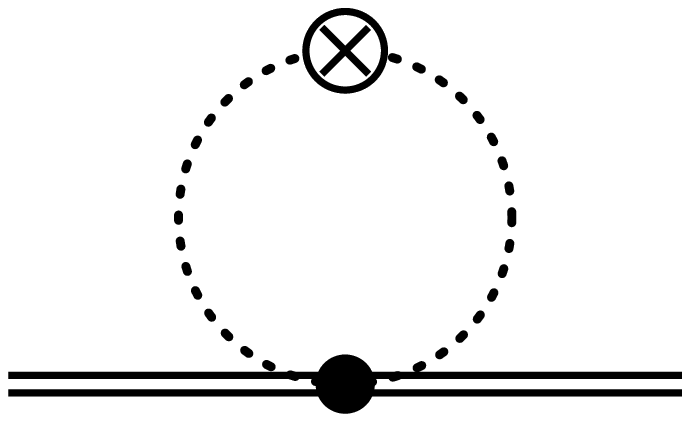} \\
		\vspace*{1.0cm}
		\includegraphics[height=1.5cm]{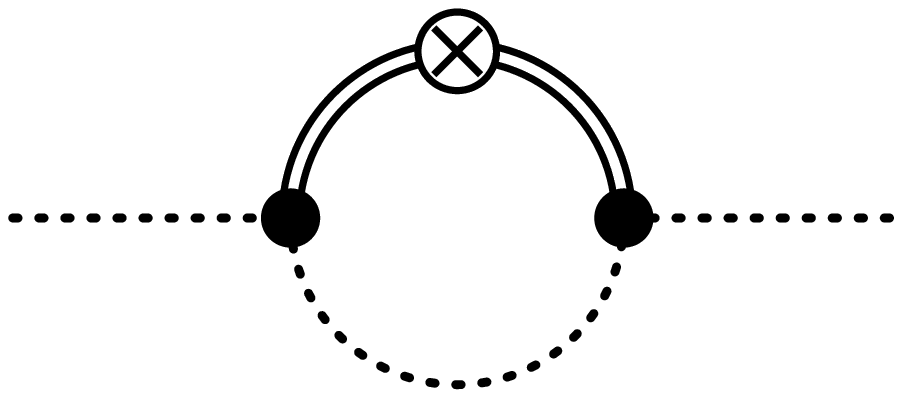} \qquad
		\includegraphics[height=1.5cm]{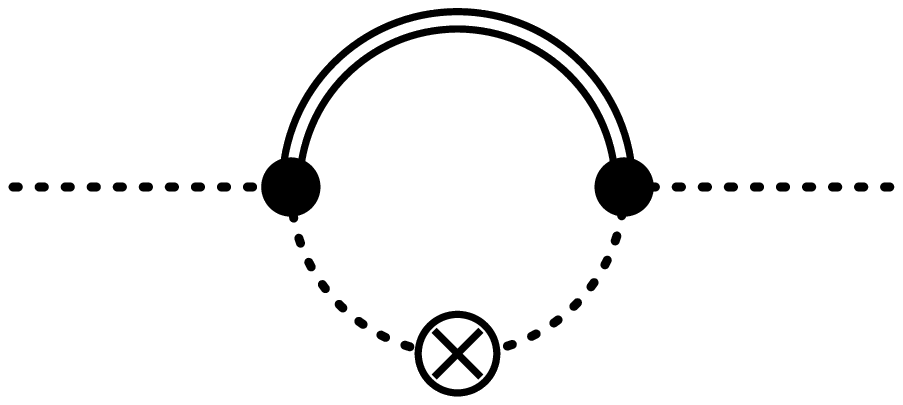} \qquad
		\includegraphics[height=1.5cm]{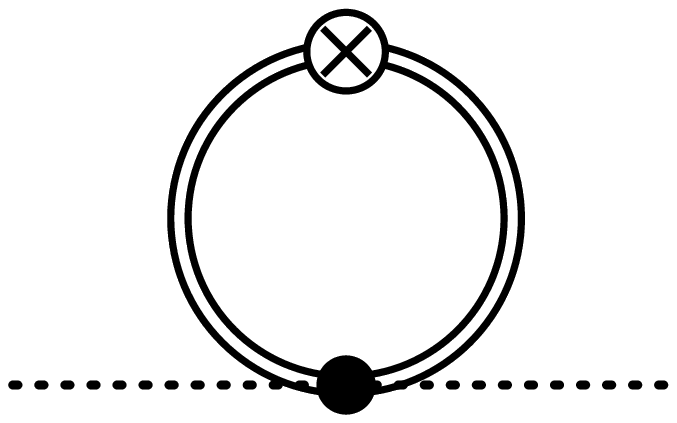}
		\caption{\footnotesize Diagrammatic representation corresponding to the r.h.s. of Eq. \eqref{Flow_Eq_2point}. The first row correspond to the flow of the graviton 2-point function $\delta^2 \Gamma_k/\delta h^2$. In the second row we include diagrams representing the flow of the ghost 2-point function $\delta^2 \Gamma_k/\delta c \delta \bar{c}$. The double-line style correspond to the graviton, while the Fadddeev-Popov ghosts are represented by dotted lines.}
		\label{Flow_Gamma_2_hh}
	\end{center}
\end{figure}

\begin{figure}[t]
	\begin{center}
		\includegraphics[height=5.8cm]{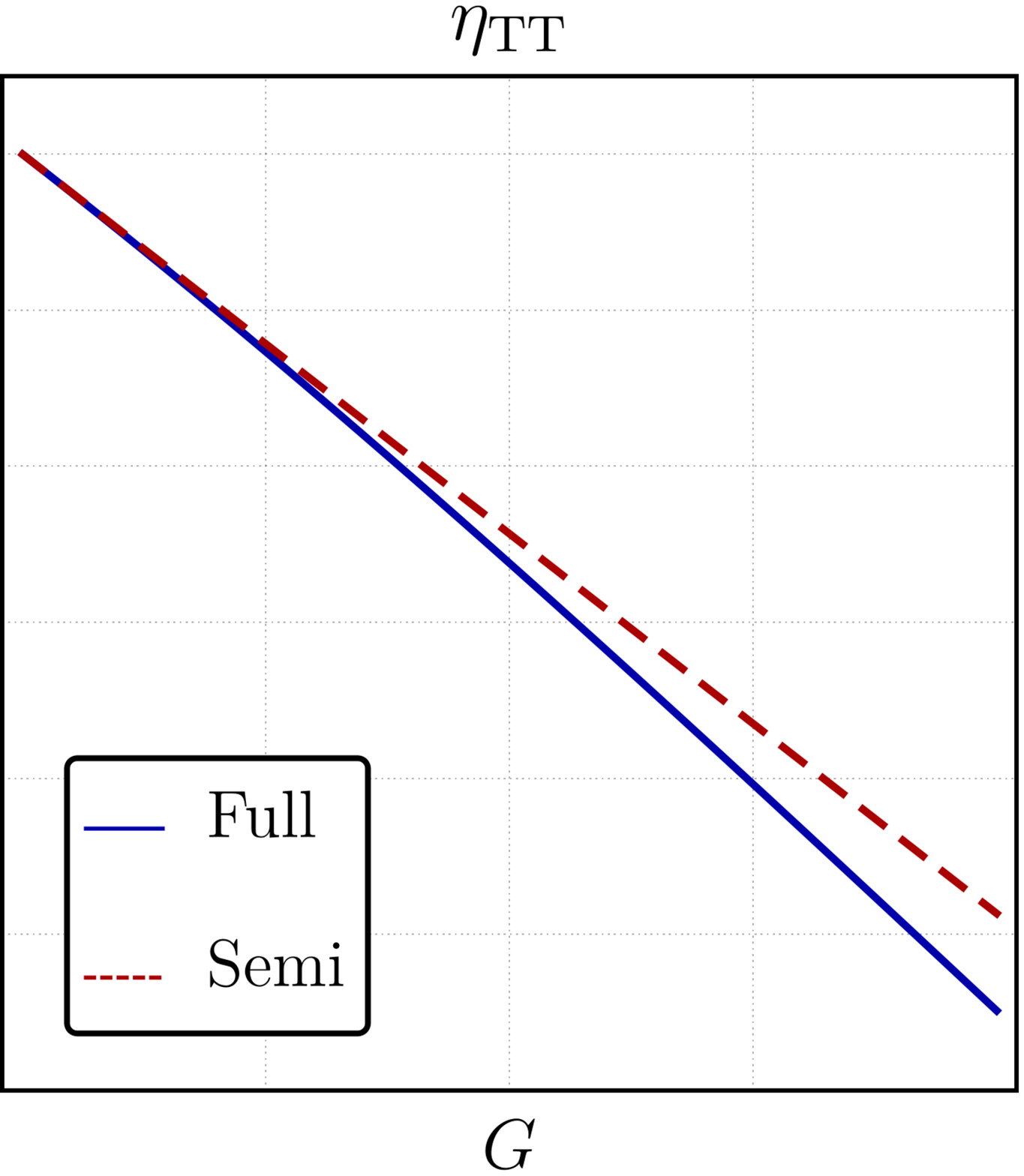} \quad
		\includegraphics[height=5.8cm]{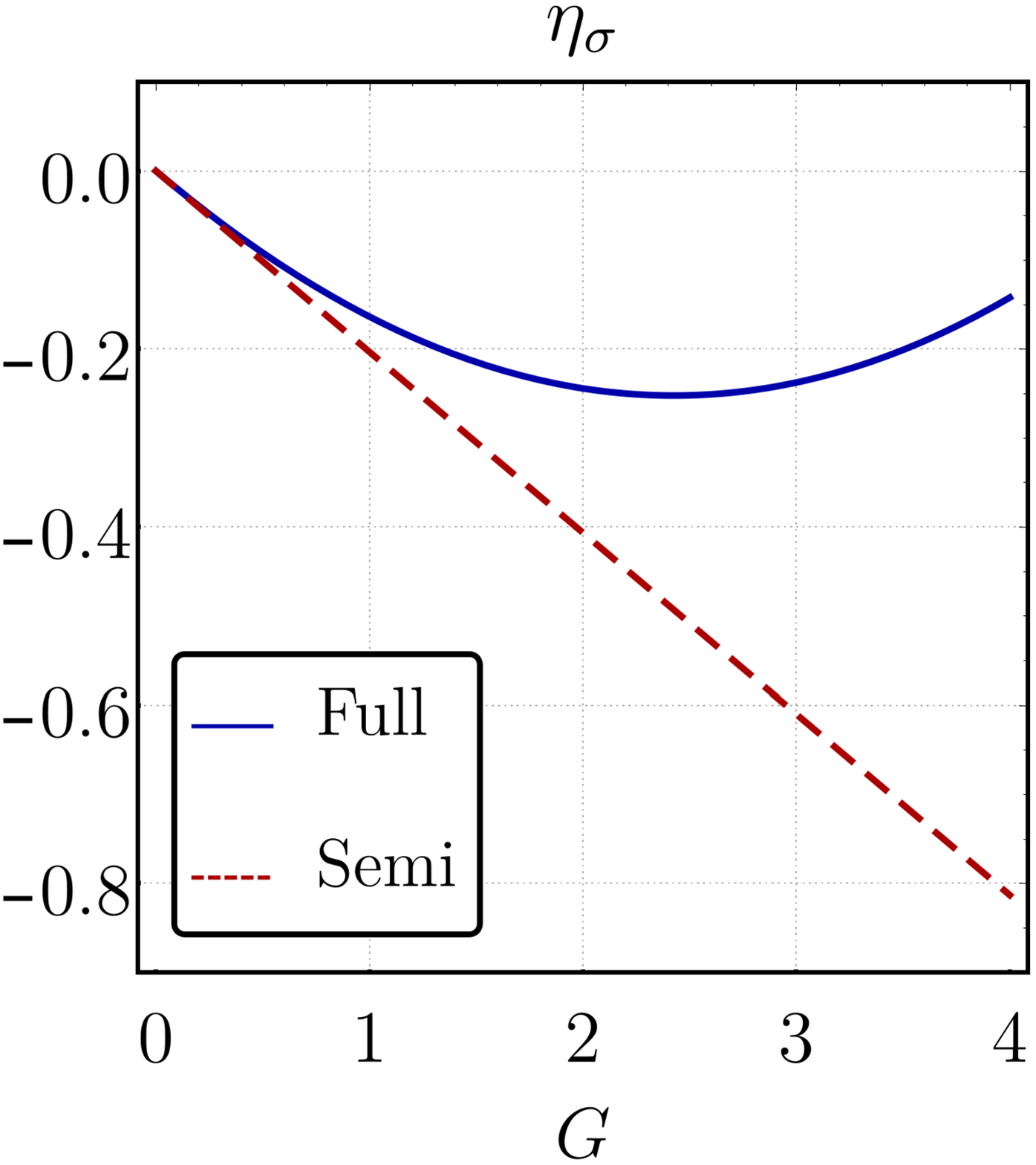} \quad
		\includegraphics[height=5.8cm]{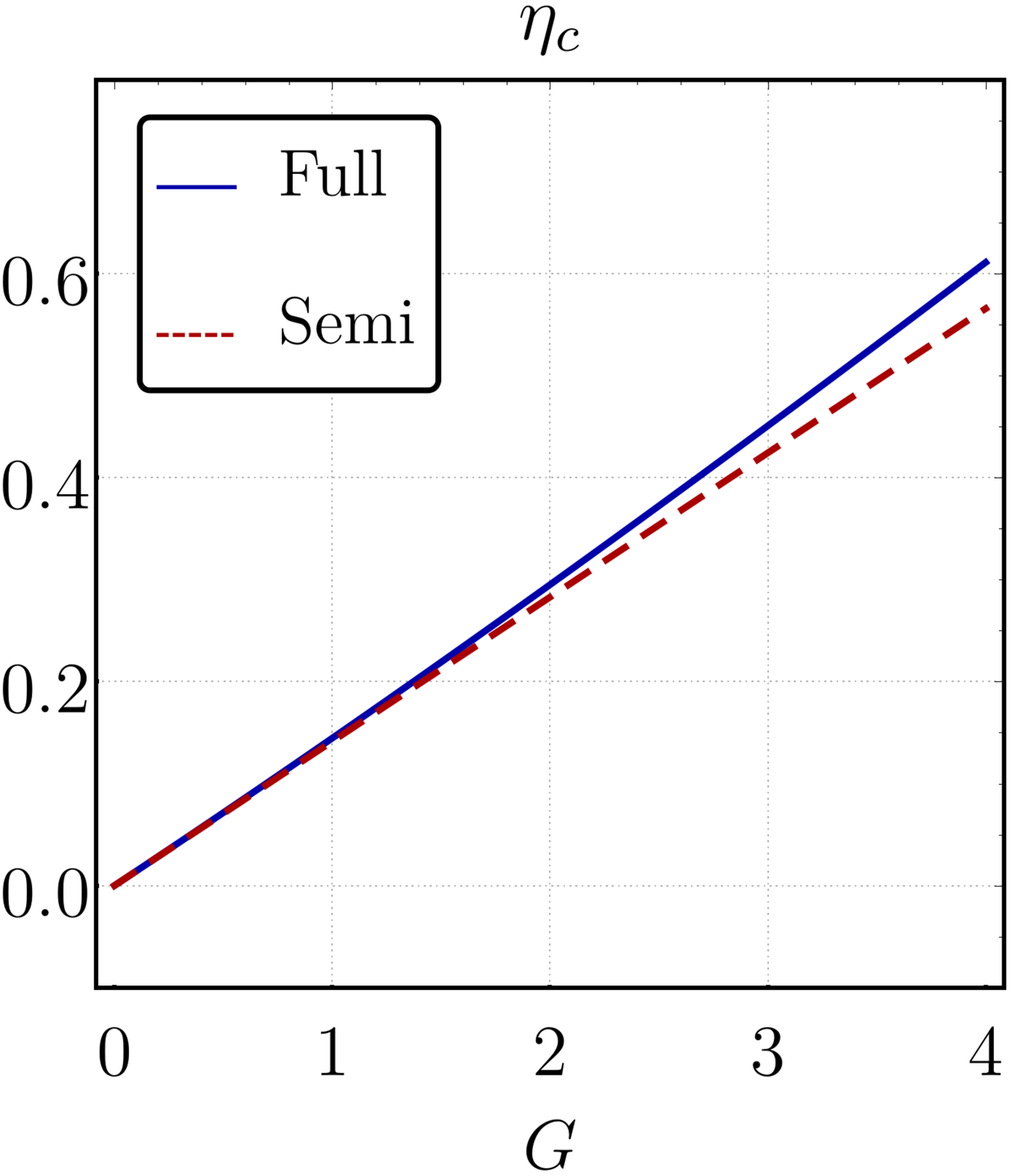} 
		\caption{\footnotesize 
		We show the anomalous dimensions $\eta_\TT$, $\eta_\sigma$ and $\eta_c$, in terms of the dimensionless Newton's coupling, in the case corresponding to $m_{k,\TT}^2 = m_{k,\sigma}^2 = 0 $ in different schemes the full result and the semi-perturbative approximation.}
		\label{eta_Zero_masses_Variable_G}
	\end{center}
\end{figure}

Starting from the simplest situation, we first set $\tilde{m}_{k,\TT}^2 = \tilde{m}_{k,\sigma}^2= 0$ in order to explore the behavior of the anomalous dimension in terms of the dimensionless Newton's coupling $G_k = k^{2} G_{k,\textmd{N}}$. In Fig. \ref{eta_Zero_masses_Variable_G}, we plot the anomalous dimensions $\eta_\TT$, $\eta_\sigma$ and $\eta_c$ as functions of $G_k$ based on two different types of results: full and semi-perturbative. The full result is obtained by solving Eqs. \eqref{eta_TT_result}, \eqref{eta_sigma_result} and \eqref{eta_Gh_result} for $\eta_\TT$, $\eta_\sigma$ and $\eta_c$ without any further approximation. The semi-perturbative calculation, on the other hand, corresponds to the anomalous dimension obtained by setting the $\eta$'s to zero on the r.h.s. of \eqref{eta_TT_result}, \eqref{eta_sigma_result} and \eqref{eta_Gh_result}. As we can observe in Fig. \ref{eta_Zero_masses_Variable_G}, for $\tilde{m}_{k,\TT}^2 = \tilde{m}_{k,\sigma}^2= 0$, the full and semi-perturbative results exhibit the same qualitative behavior both in the case of $\eta_\TT$ and $\eta_c$. Nevertheless, the situation is different in the case of $\eta_\sigma$. In this case, the full result exhibit a considerable deviation from the linear behavior corresponding to the semi-perturbative approximation.

With inclusion of the masses $m_{k,\TT}^2$ and $m_{k,\sigma}^2$, we investigate the viability of a UV completion within the extended truncation which includes symmetry-breaking terms. This means that we look for fixed point solutions of the partial system of RG equations
\begin{subequations}
	\begin{align}\label{Partial_RG_1}
	\pt_t \tilde{m}_{k,\TT}^2 = -(2-\eta_\TT) \,\tilde{m}_{k,\TT}^2 + f_\TT(\tilde{m}_{k,\TT}^2,\tilde{m}_{k,\sigma}^2,\eta_\TT,\eta_\sigma,\eta_c,G_k) \,,
	\end{align}
	\begin{align}\label{Partial_RG_2}
	\pt_t \tilde{m}_{k,\sigma}^2 = -(2-\eta_\sigma) \,\tilde{m}_{k,\sigma}^2 + f_\sigma(\tilde{m}_{k,\TT}^2,\tilde{m}_{k,\sigma}^2,\eta_\TT,\eta_\sigma,\eta_c,G_k) \,.
	\end{align}
\end{subequations}
The explicit form of the functions $f_\TT$ and $f_\sigma$ can be read off from Eqs. \eqref{mass_TT} and \eqref{mass_sigma}. It is interesting to emphasize that both $f_\TT$ and $f_\sigma$ are non-vanishing for $G_k \neq 0$, which confirms that even if we set $\tilde{m}_{k,\TT}^2 = \tilde{m}_{k,\sigma}^2 = 0$ at some RG-scale $k$, symmetry-breaking mass terms would be generated due to graviton self-interactions. The partial system corresponding to Eqs. \eqref{Partial_RG_1} and \eqref{Partial_RG_2} is not closed, since, at this level, the Newton's coupling appears as an external parameter. Within this setting, we perform the search of fixed points solutions for $\pt_t \tilde{m}_{k,\TT}^2 = 0$ and $\pt_t \tilde{m}_{k,\sigma}^2 = 0$, by assuming the existence of a fixed point for the Newton coupling and treating its value as a free parameter. This strategy allows us to explore properties of the $m_{k,i}^2$'s and $\eta_i$'s without relying on some specific expression (computed within an approximation scheme) for the running of the Newton's coupling. 

\begin{figure}[t]
	\begin{center}
		\includegraphics[height=5.8cm]{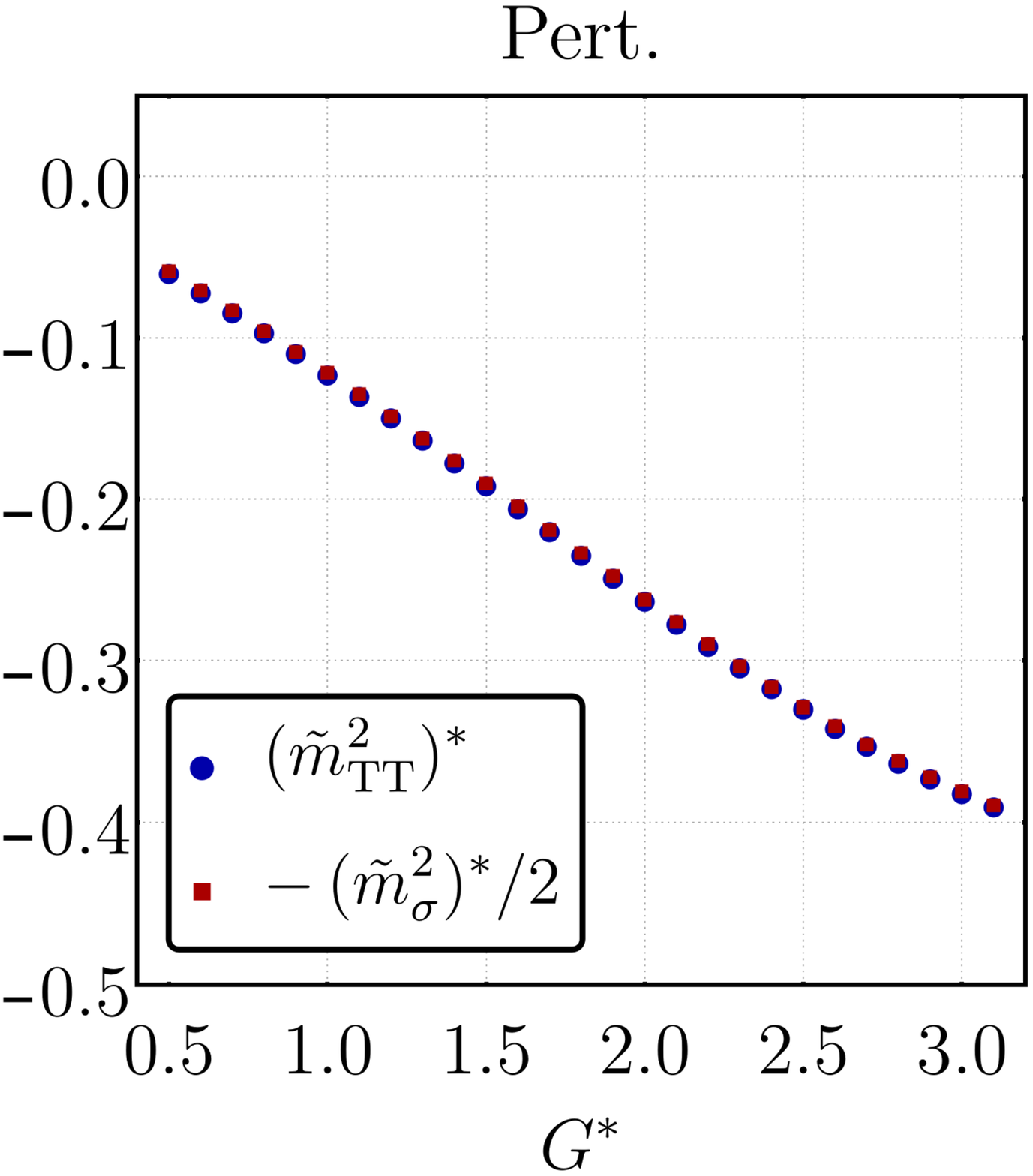} \quad
		\includegraphics[height=5.8cm]{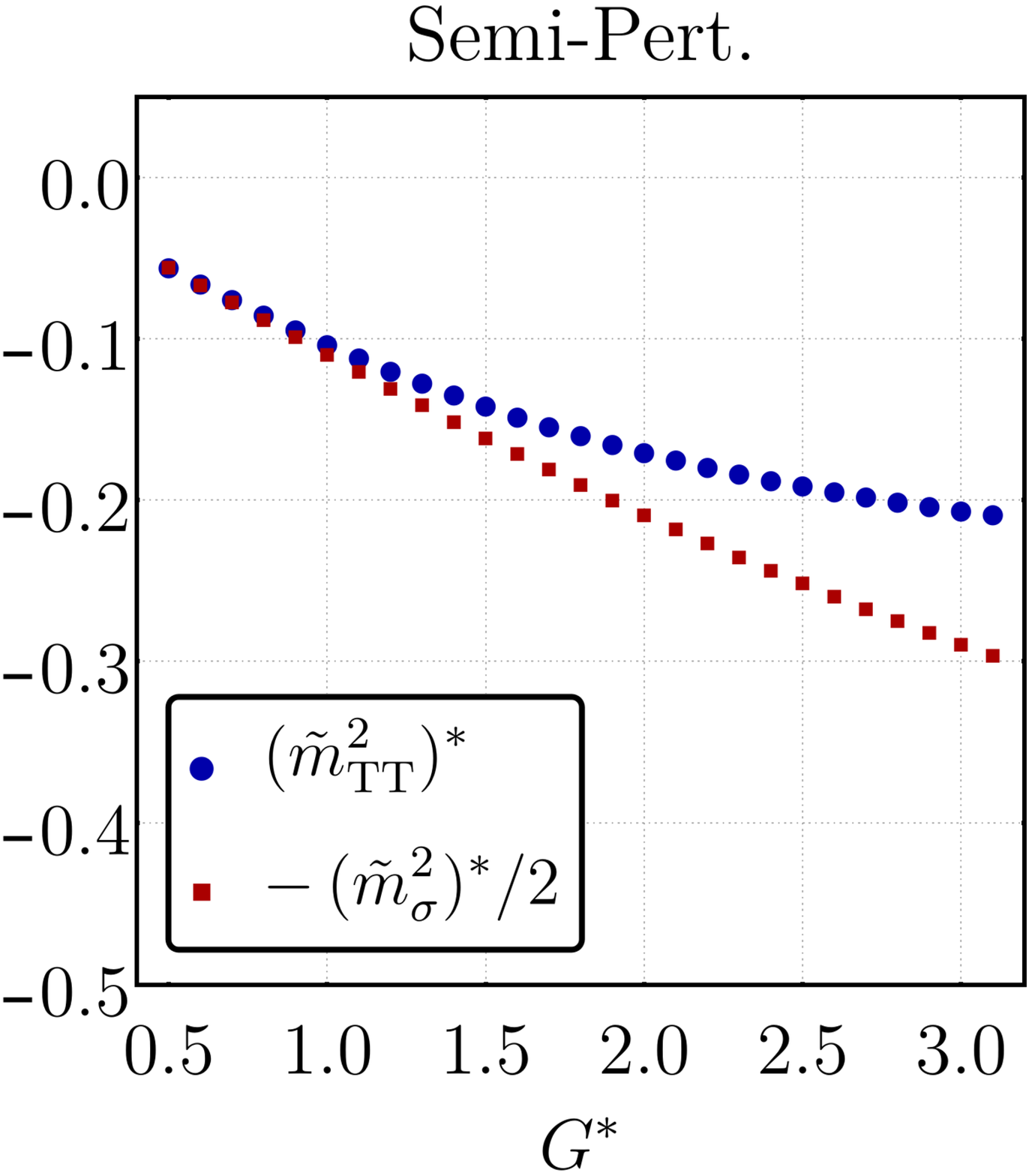} \quad
		\includegraphics[height=5.8cm]{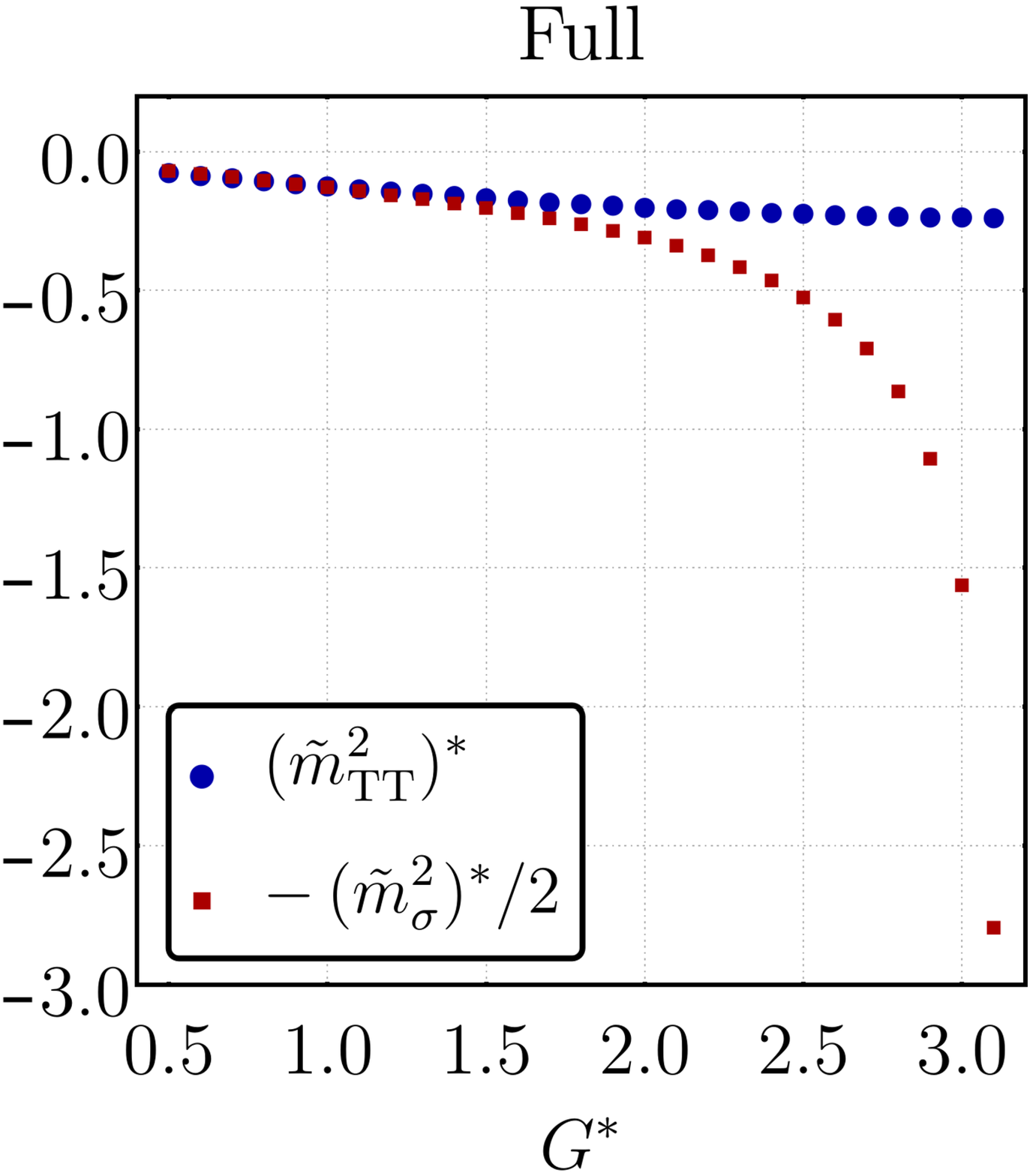} 
		\caption{\footnotesize 
		Fixed point solutions for the partial system composed by Eqs. \eqref{Partial_RG_1} and $\eqref{Partial_RG_2}$. In this case, the fixed point value for the dimensionless Newton's coupling, $G^*$, appears as an external variable.}
		\label{FP_masses_Variable_G}
	\end{center}
\end{figure}

In Fig. \ref{FP_masses_Variable_G} we plot the fixed point values $(\tilde{m}_{\TT}^2)^*$ and $-(\tilde{m}_{\sigma}^2)^*/2$ as functions of $G^*$. For the sake of comparison, we have considered three different schemes. The perturbative approximation is obtained by setting the $\eta$'s to zero in Eqs. \eqref{Partial_RG_1} and \eqref{Partial_RG_2}. In this case, both $(\tilde{m}_{\TT}^2)^*$ and $-(\tilde{m}_{\sigma}^2)^*/2$ exhibit the same values along the range under consideration. The semi-perturbative regime is defined by setting the anomalous dimensions to zero in the functions $f_\TT$ and $f_\sigma$, but using the semi-perturbative expressions for $\eta_\TT$ and $\eta_\sigma$ in the first term on the r.h.s. of Eqs.\eqref{Partial_RG_1} and \eqref{Partial_RG_2}. Within this approximation we note that the separation between the fixed point values of $(\tilde{m}_{\TT}^2)^*$ and $-(\tilde{m}_{\sigma}^2)^*/2$ increases with $G^*$. Finally, the full result correspond to fixed point solutions of Eqs.\eqref{Partial_RG_1} and \eqref{Partial_RG_2} without further approximations. As one can observe, the fixed point value of $-(\tilde{m}_{\sigma}^2)^*/2$ hits a pole around $G^* = 3.2$. On the other hand, $(\tilde{m}_{\TT}^2)^*$ displays a small variation along the range considered for $G^*$. Furthermore, we note that $(\tilde{m}_{\TT}^2)^*$ and $-(\tilde{m}_{\sigma}^2)^*/2$ approach the coincident values when $G^*$ is small ($G^*\lesssim1.5$).

\begin{figure}[t]
	\begin{center}
		\includegraphics[height=5.8cm]{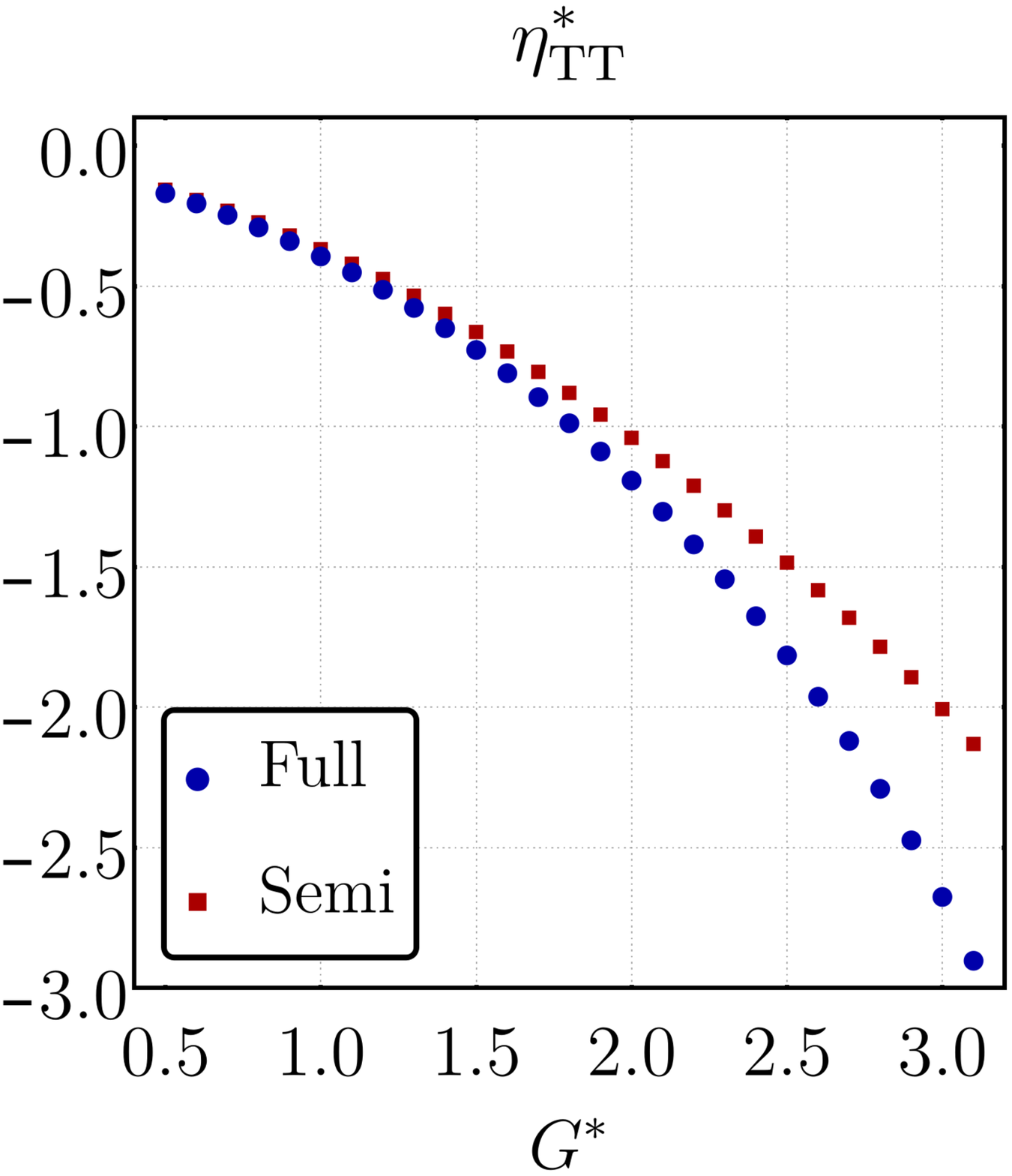} \quad
		\includegraphics[height=5.8cm]{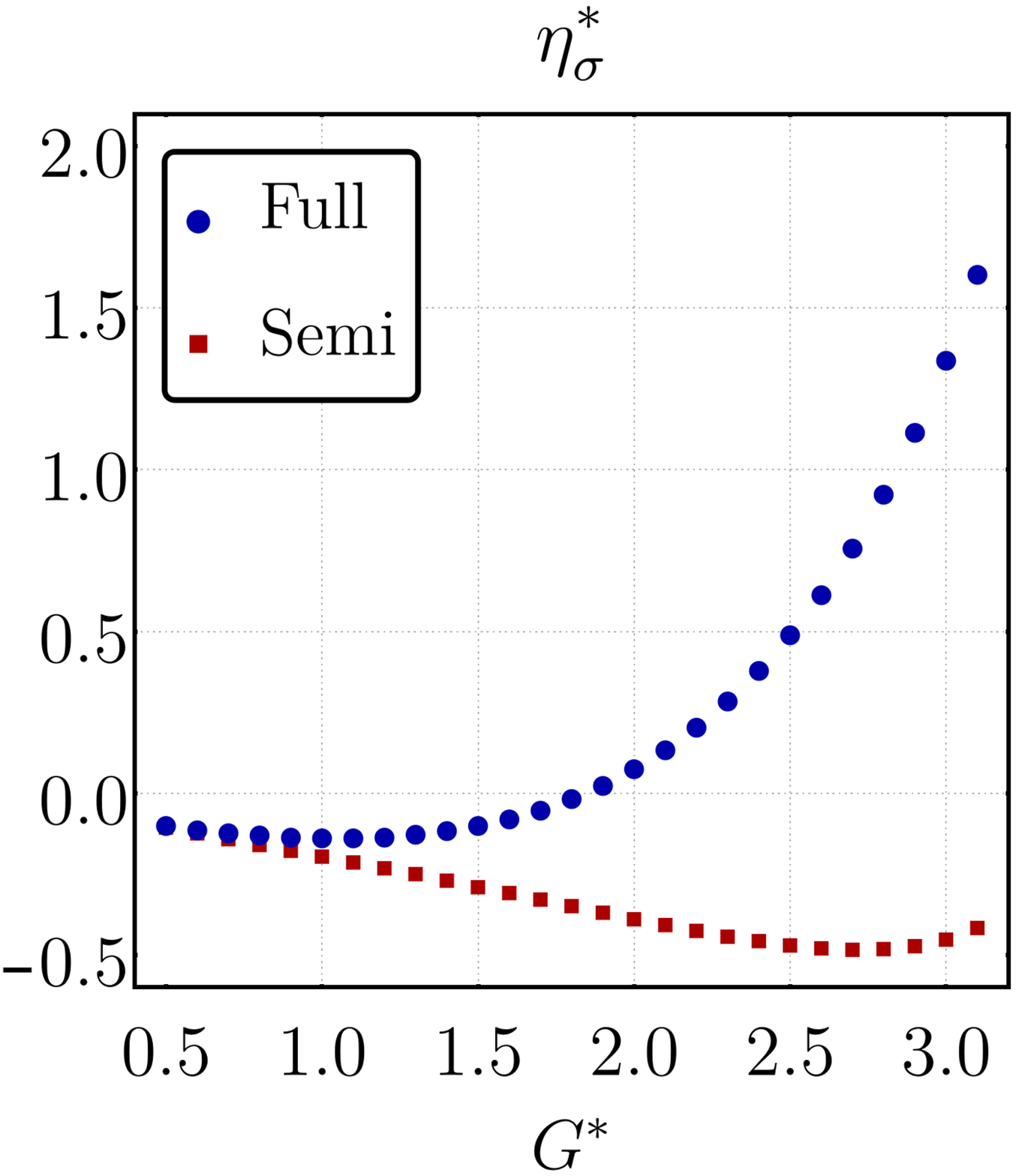} \quad
		\includegraphics[height=5.8cm]{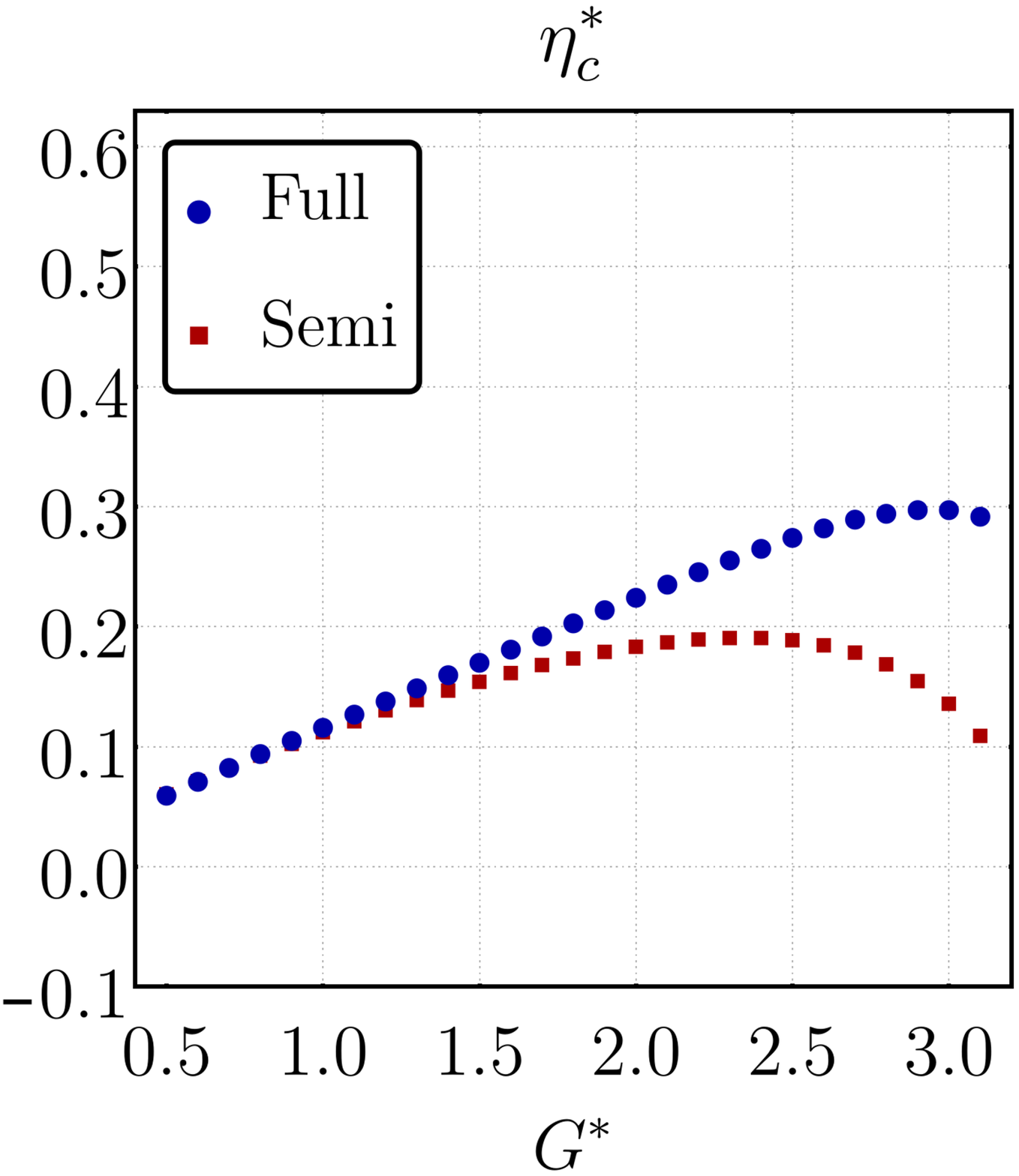} 
		\caption{\footnotesize 
		Anomalous dimensions $\eta_\TT$, $\eta_\sigma$ and $\eta_c$ evaluated at the fixed point solutions depicted in Fig. \ref{FP_masses_Variable_G}.}
		\label{eta_Variable_G}
	\end{center}
\end{figure}

In Fig. \ref{eta_Variable_G} we show the anomalous dimensions $\eta_\TT$, $\eta_\sigma$ and $\eta_c$ evaluated at the fixed point solutions of Eqs.\eqref{Partial_RG_1} and \eqref{Partial_RG_2}. Comparing Figs. \ref{eta_Zero_masses_Variable_G} and \ref{eta_Variable_G}, we observe clear differences with respect to the behavior of the anomalous dimensions with and without setting the masses to zero. In particular, we note that the absolute values of $\eta_\TT^*$ and $\eta_\sigma^*$ grow faster in the case where we take into account the mass parameters, signaling that non-perturbative effects may become relevant for smaller values of $G^*$ in comparison with the case $\tilde{m}_{k,i}^2 = 0$. In addition, one can observe that $\eta_\sigma^*$ becomes larger than $2$ around $G^*\sim 3.2$. In connection to this point, it has been argued that for a class of regulators results with $\eta^*>2$ become unreliable \cite{Meibohm:2015twa}. Since the regulator used in this paper belongs to this class, one can argue that internal consistency for our results (with $\tilde{m}_{k,i}^2 \neq 0$) requires $G^*$ to be smaller than $3.2$.

\section{Renormalization Group Flow and Fixed Point Structure \label{RGFlowAndFP}}

Up to this point we have considered the fixed point values of the Newton's coupling as a free parameter. In this section we explore the fixed point structure including explicit results for the beta function of the dimensionless coupling $G_k$. The running of $G_k$ can be computed in several ways, each one corresponding to a different ``avatar'' of the Newton's coupling (see \cite{Eichhorn:2018akn,Eichhorn:2018ydy} for a detailed discussion). The relation between different avatars are encoded in the modified Slavnov-Taylor identities (mSTI's) and modified Nielsen identities (mNI's) \cite{Eichhorn:2018akn,Eichhorn:2018ydy}. In this paper, we extract the running of $G_k$ from the flow of $\Gamma_k[\varphi=0;\bar{g}]$. In order to simplify the computations, the background is considered to be a 4-sphere. In such a case, the running of the dimensionless Newton's coupling can be obtained by the following expression
\begin{align}
\pt_t G_k = 2\,G_k + 16\pi \,k^{-2}\,G_k^2 \,\times
\bigg[ \frac{\pt }{\pt \bar{R}} \bigg(\frac{\pt_t \Gamma_k}{V(S^4)} \bigg) \bigg]_
{\substack{\varphi=0,\,\bar{R}=0}},
\end{align}
where $V(S^4)$ stands for the volume of the 4-sphere.
Using standard heat-kernel methods (see, e.g., \cite{Percacci:2017fkn}) in order to compute the trace in the r.h.s. of \eqref{Flow_Eq} leads to
\begin{align}\label{Beta_G_UG}
\pt_t G_k = 2 G_k + \frac{G_k^2}{24\pi} \Big( \mathcal{A}(\tilde{m}_{k,\TT}^2,\tilde{m}_{k,\sigma}^2) + \mathcal{B}_{\TT}(\tilde{m}_{k,\TT}^2) \,\eta_\TT 
+ \mathcal{B}_\sigma(\tilde{m}_{k,\sigma}^2) \,\eta_\sigma + \mathcal{B}_{c} \,\eta_c \Big) \,.
\end{align}
The coefficients $\mathcal{A}(\tilde{m}_{k,\TT}^2,\tilde{m}_{k,\sigma}^2)$, $\mathcal{B}_{\TT}(\tilde{m}_{k,\TT}^2)$, $\mathcal{B}_\sigma(\tilde{m}_{k,\sigma}^2)$ and $\mathcal{B}_{c}$ are scheme dependent quantities that can be computed within the truncation defined in Sect. \ref{sect:setup}. In table \ref{Table_Coeffs} we present the explicit results for these coefficients in terms of two types of regularization schemes distinguished by the choice of coarse-graining operators, namely, using Bochner (type I) and Lichnerowicz (type II) Laplacians \cite{Codello:2008vh}. It is important to remark that $\pt_t G_k$ involves (via anomalous dimension contributions) the avatars of the Newton's coupling extracted from 3- and 4-graviton vertices and graviton-ghost vertices. We take as an additional approximation the identification of all these avatars with a single coupling $G_k$. 
\begin{center}
	\begin{table}[t]
		\begin{tabular}{|c|c|c|c|c|}
			\hline\hline
			$\qquad\qquad\quad$& $\mathcal{A}(\tilde{m}_{k,\TT}^2,\tilde{m}_{k,\sigma}^2)$ & $\mathcal{B}_\TT(\tilde{m}_{k,\TT}^2)$ & $\mathcal{B}_\sigma(\tilde{m}_{k,\sigma}^2)$ & $\mathcal{B}_c$ \\
			\hline
			%%%%%%%%%%%%%%%%%%%%%%%%%%%%%%%%%%%%%%%%%%%%%%%%%%%%%%%%%%%%%%%%%%%%%%%%%%%%%%%%%%%%%%%%%%%%%%%%%%
			Type I & $-\frac{ 30\,(3+2\,\tilde{m}_{k,\TT}^2)}{3\,(1+\tilde{m}_{k,\TT}^2)^2}  
			+\frac{4}{1+\tilde{m}_{k,\sigma}^2} - 19$  & 
			$\quad$ $\frac{ 5\,(4 + 3 \,\tilde{m}_{k,\TT}^2)}{3\,(1+\tilde{m}_{k,\TT}^2)^2}$ $\quad$ & 
			$\quad$ $-\frac{1}{1+\tilde{m}_{k,\sigma}^2}$ $\quad$ & $\qquad$ 6 $\qquad$ \\
			%%%%%%%%%%%%%%%%%%%%%%%%%%%%%%%%%%%%%%%%%%%%%%%%%%%%%%%%%%%%%%%%%%%%%%%%%%%%%%%%%%%%%%%%%%%%%%%%%%
			Type II & $-\frac{10\,(7+10\, \tilde{m}_{k,\TT}^2)}{(1+\tilde{m}_{k,\TT}^2)^2}  
			+\frac{4}{1+\tilde{m}_{k,\sigma}^2}  - 10$ & 
			$\frac{	5\,(4 + 5 \,\tilde{m}_{k,\TT}^2)}{(1+\tilde{m}_{k,\TT}^2)^2}$ & 
			$-\frac{1}{1+\tilde{m}_{k,\sigma}^2}$ & 0 \\
			\hline\hline
		\end{tabular}
	\caption{\footnotesize Explicit coefficients $\mathcal{A}(\tilde{m}_\TT^2,\tilde{m}_\sigma^2)$, $\mathcal{B}_\TT(\tilde{m}_\TT^2)$, $\mathcal{B}_\sigma(\tilde{m}_\sigma^2)$ and $\mathcal{B}_{c}$ for two types of coarse-graining operators. Here, we use the nomenclature ``type I'' to designate the case where the coarse-graining operator corresponds to the Bochner-Laplacian $\Delta_\textmd{B} = -\bar{\nabla}^2$. The nomenclature ``type II'' corresponds to the choice of Lichnerowicz-Laplacian defined as $\Delta_\textmd{L} = -\bar{\nabla}^2 + \alpha \bar{R}$ (with $\alpha = 2/3$, $\alpha = 1/4$ and $\alpha = 0$, respectively, for transverse-traceless tensors, transverse vectors and scalars) on spherical backgrounds.}
	\label{Table_Coeffs}
	\end{table}
\end{center}

At this point we should emphasize the difference of the investigation performed here with respect to previous results in asymptotically safe unimodular gravity \cite{Eichhorn:2013xr,Eichhorn:2015bna,Benedetti:2015zsw}. The main difference lies on the fact that previous computations in this setting were done within the background approximation. In such a case, the 2-point function $\delta^2 \Gamma_k/\delta h^2|_{\varphi=0}$ is identified with $\delta^2 \Gamma_k/\delta \bar{g}^2|_{\varphi=0}$ (plus gauge-fixing contributions). In this case, the closure of the flow equation for the Newton's coupling is obtained with a ``RG-improved'' (see, e.g., \cite{Codello:2013fpa}) anomalous dimensions $\eta_\TT = \eta_\sigma = -2+G_{k}^{-1} \pt_t G_{k}$ and $\eta_c = 0$. Furthermore, the background approximation does not include the regulator induced masses $m_{k,\TT}^2$ and $m_{k,\sigma}^2$. In this paper, we perform two steps beyond the background approximation: i) we have computed the anomalous dimensions $\eta_\TT$, $\eta_\sigma$ and $\eta_c$ using a derivative expansion; ii) the truncation considered here includes symmetry deformation effects parameterized by the masses $m_{k,\TT}^2$ and $m_{k,\sigma}^2$. Another important difference in comparison with previous investigations in asymptotically safe unimodular gravity is the inclusion of an extra trace in the flow equation (see Eq. \eqref{flowequnim}) accounting for an appropriate treatment of the path integral measure for the gauge-fixing of the \textit{TDiff} invariance.

\begin{figure}[t]
	\begin{center}
		\includegraphics[height=5.8cm]{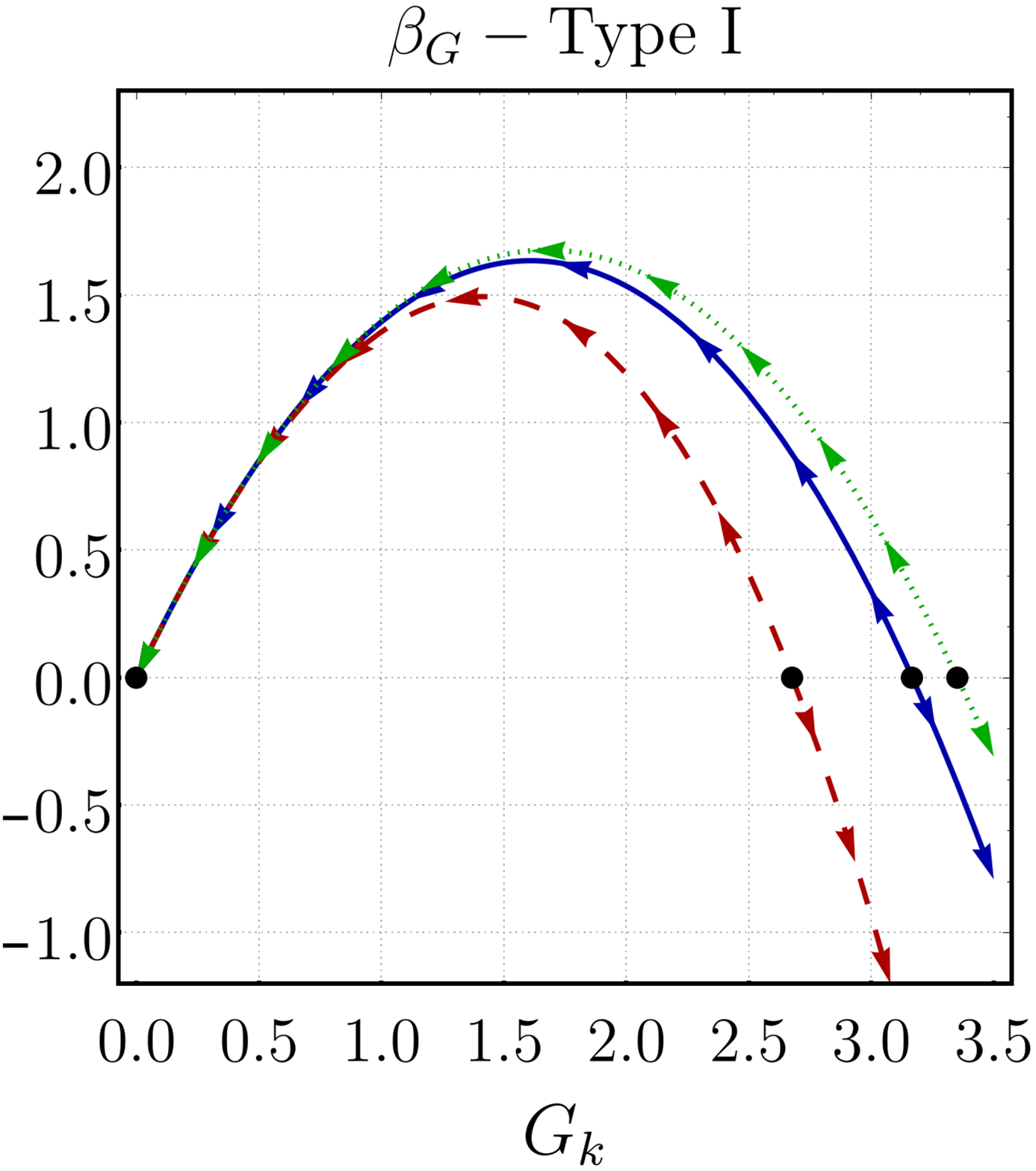} \quad\qquad
		\includegraphics[height=5.8cm]{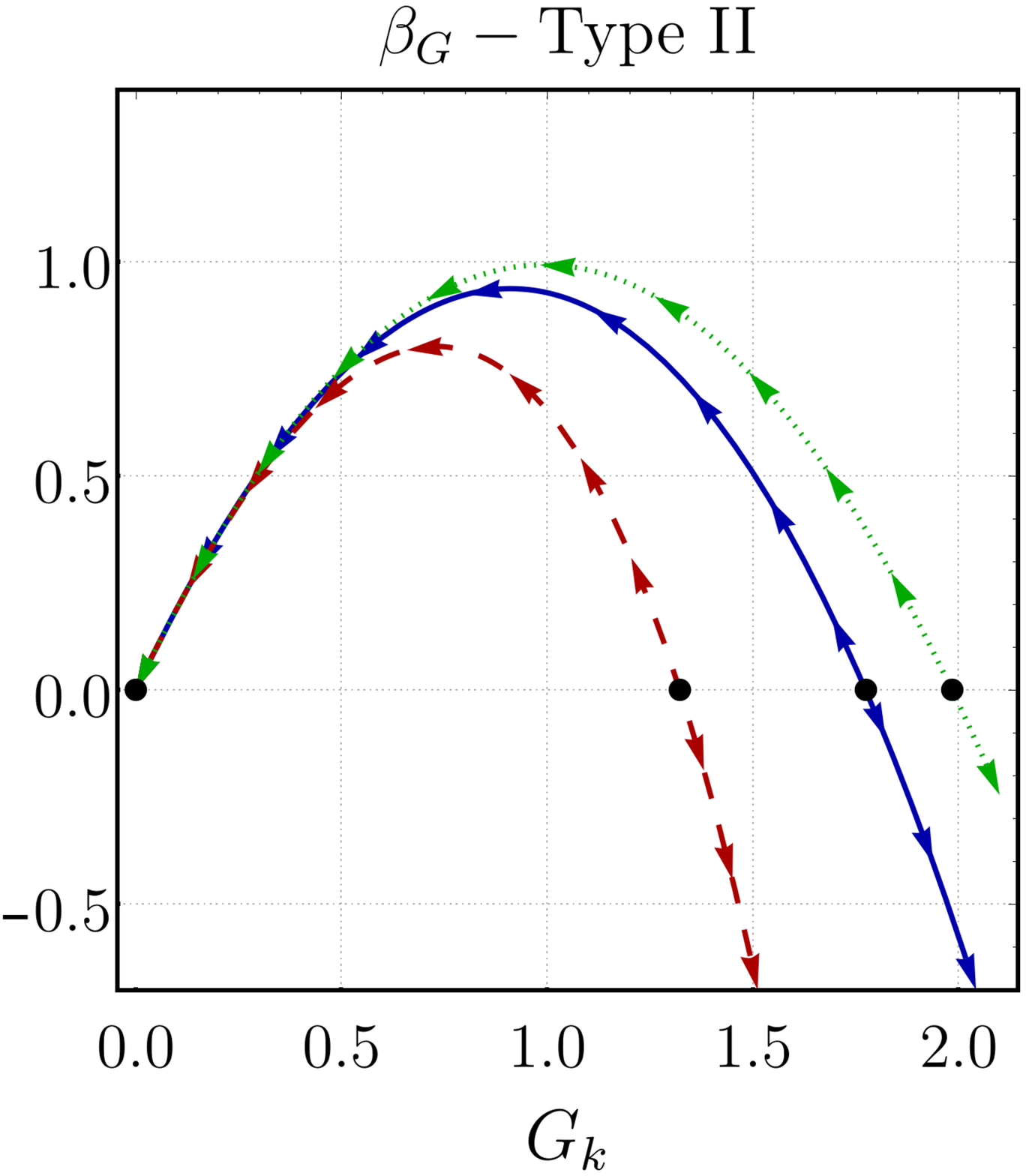}  
		\caption{\footnotesize Beta function for the dimensionless Newton's coupling in the case $\tilde{m}_\TT^2=\tilde{m}_\sigma^2 = 0$. The plot on the left (right) correspond to the type I (II) coarse-graining operator. The blue (continuous) line corresponds to the case where the anomalous dimensions were replaced by the results reported in the App. \ref{Explicit}. The red (dashed) and green (dotted) lines represent ``RG improved'' and 1-loop closure, respectively. Conventionally, the arrows point towards the infrared.}
		\label{Beta_G_Zero_Masses}
	\end{center}
\end{figure}

In view of a better understanding concerning the impact of the anomalous dimensions and the symmetry breaking mass parameters, it is useful to consider different approximations. Let us start with the case where the masses $m_{k,\TT}^2$ and $m_{k,\sigma}^2$ are set to zero along flow. In this case the system of RG equations reduces to $\pt_t G_k = \beta_G(G_k)$, where the function $\beta_G(G_k)$ corresponds to the r.h.s. of \eqref{Beta_G_UG} with $\tilde{m}_{k,\TT}^2=\tilde{m}_{k,\sigma}^2 = 0$ and using the anomalous dimensions $\eta_\TT$, $\eta_\sigma$ and $\eta_c$ reported in App. \ref{Explicit}. In Fig. \ref{Beta_G_Zero_Masses} we show the beta function $\beta_G(G_k)$ for the two types of regularization schemes considered here. For the sake of comparison we also include the 1-loop and the ``RG improved'' closure where, instead of using the anomalous dimensions reported in the App. \ref{Explicit}, we use the prescriptions $\eta_\TT = \eta_\sigma = \eta_c = 0$ and $\eta_\TT = \eta_\sigma = -2+G_{k}^{-1} \pt_t G_{k}$ and $\eta_c = 0$, respectively. In all cases we observe an UV attractive interacting fixed point for the dimensionless Newton's coupling. The numerical values for the fixed points and critical exponents (see table \ref{Table_Fixed_Points_ZeroMasses}) are, as usual, scheme and approximation dependent. However, by looking at Fig. \ref{Beta_G_Zero_Masses} we observe the same qualitative features in all the considered approximations.

\begin{center}
	\begin{table}[t]
		\begin{tabular}{|c|c|c|c|c|c|c|}
			\hline\hline
			                              & $\qquad G^* \quad$ & $\quad\,\,\,$ $\theta$ $\quad\,\,\,$ & $\quad$ $\eta_\TT^*$ $\quad$ & $\quad$ $\eta_\sigma^*$ $\quad$ & $\quad$ $\eta_c^*$ $\quad$ \\
			\hline                              
			1-Loop  -- Type I             & $3.35$            & $2$                    & $0$                          & $0$                             & $0$                        \\
			1-Loop  -- Type II            & $1.98$            & $2$                    & $0$                          & $0$                             & $0$                        \\
			``RG-Improv.'' -- Type I      & $2.67$            & $2.50$                   & $-2$                         & $-2$                            & $0$                        \\
			``RG-Improv.'' -- Type II     & $1.32$            & $3.00$                   & $-2$                         & $-2$                            & $0$                        \\
			$\eta$'s from D.E. -- Type I  & $3.18$            & $2.14$                   & $-0.86$                      & $-0.23$                         & $0.48$                     \\
			$\quad$ $\eta$'s from D.E. -- Type II $\quad$& $1.77$            & $2.23$                   & $-0.46$                      & $-0.23$                         & $0.26$                    \\
			\hline\hline
		\end{tabular}
		\caption{\footnotesize Fixed point structure associated with the case $\tilde{m}_{k,\TT}^2=\tilde{m}_{k,\sigma}^2 = 0$. Here we report results obtained using both types of coarse-graining operators and for the different approximations concerning the anomalous dimensions.}
		\label{Table_Fixed_Points_ZeroMasses}
	\end{table}
\end{center}

The approximation $\tilde{m}_{k,\TT}^2=\tilde{m}_{k,\sigma}^2 = 0$ is not self-consistent, since, as it was mentioned in the previous section, even if not included in the original truncation, the symmetry-breaking masses are generated by the RG-flow. Here we consider the full system describing the RG flow in the truncated theory space characterized by $G_k$, $\tilde{m}_{k,\TT}^2$ and $\tilde{m}_{k,\sigma}^2$. In table \eqref{Table_Fixed_Points} we summarize our findings for the fixed point structure. The corresponding critical exponents are shown in table \ref{Table_Crit_Exp}. Confronting the results exhibited in table \ref{Table_Fixed_Points_ZeroMasses} and \ref{Table_Fixed_Points} we observe that the inclusion of symmetry breaking masses shift $G^*$ towards smaller values in comparison with the case $\tilde{m}_{k,\TT}^2=\tilde{m}_{k,\sigma}^2 = 0$. In the case of the 1-loop approximation, we observe fixed point values with $(\tilde{m}^2_{\TT})^* \approx - \frac{1}{2}(\tilde{m}^2_{\sigma})^*$, in accordance with the analysis discussed in the previous section (see Fig. \ref{FP_masses_Variable_G}). Within the full closure the fixed point values for $(\tilde{m}^2_{\TT})^*$ and $- \frac{1}{2}(\tilde{m}^2_{\sigma})^*$ exhibit a considerable difference, in special, in the type I regularization scheme. We also observe substantial differences concerning the fixed point values for the different avatars of the graviton anomalous dimensions, $\eta_\TT^*$ and $\eta_\sigma^*$. In particular, we note that $\eta_\sigma^*$ changes the sign according to the type of the coarse-graining operator.

\begin{center}
	\begin{table}[t]
		\begin{tabular}{|c|c|c|c|c|c|c|}
			\hline\hline
			& $\quad$ $G^*$ $\quad$ & $\,\,$ $(\tilde{m}^2_{\TT})^*$ $\,\,$ &  $-(\tilde{m}^2_{\sigma})^*/2$ $\,$  & $\quad$ $\eta_\TT^*$ $\quad$ & $\quad$ $\eta_\sigma^*$ $\quad$ & $\quad$ $\eta_c^*$ $\quad$ \\
			\hline
			%%%%%%%%%%%%%%%%%%%%%%%%%%%%%%%%%%%%%%%%%%%%%%%%%%%%%%%%%%%%%%%%%%%%%%%%%%%%%%%%%%%%%%%%%%%%%%%%%%
			$\quad$ 1-Loop \& $m_{k,i}^2\neq0$ -- Type I $\quad$  & $2.30$ & $-0.30$                  & $-0.30$                     & --           & --              & --         \\
			1-Loop \& $m_{k,i}^2\neq0$ -- Type II         & $1.75$ & $-0.22$                  & $-0.22$                     & --           & --              & --         \\
			%%%%%%%%%%%%%%%%%%%%%%%%%%%%%%%%%%%%%%%%%%%%%%%%%%%%%%%%%%%%%%%%%%%%%%%%%%%%%%%%%%%%%%%%%%%%%%%%%%
			Full Closure -- Type I                        & $2.23$ & $-0.19$                  & $-0.37$                     & $-1.43$     & $0.24$         & $0.25$    \\
			Full Closure -- Type II                        & $1.52$ & $-0.15$                  & $-0.19$                     & $-0.72$     & $-0.08$        & $0.18$   \\
			\hline\hline
		\end{tabular}
	\caption{\footnotesize Fixed point structure in the truncated theory space defined by $G_k$, $\tilde{m}_{k,\TT}^2$ and $\tilde{m}_{k,\sigma}^2$. We report on results obtained using both types of coarse-graining operators. The ``1-Loop'' closure corresponds to the case where we set $\eta_\TT=\eta_\sigma=\eta_c=0$. We use the nomenclature ``Full Closure'' for the fixed point solutions involving anomalous dimensions reported in App. \ref{Explicit}.}
	\label{Table_Fixed_Points}
	\end{table}
\end{center}

The critical exponents reported in table \ref{Table_Crit_Exp} provide indications that the three couplings under investigation, $G_k$, $\tilde{m}_{k,\TT}^2$ and $\tilde{m}_{k,\sigma}^2$, are associated to UV relevant directions. At a first sight, this result suggests that the symmetry breaking masses also requires initial conditions determined by experimental observations. However, $\tilde{m}_{k,\TT}^2$ and $\tilde{m}_{k,\sigma}^2$ appear as technical artifacts as a consequence of the method used to implement the Wilsonian renormalization and do not feature any direct physical meaning. Therefore, we should not expect initial conditions on $\tilde{m}_{k,\TT}^2$ and $\tilde{m}_{k,\sigma}^2$ coming from  ``experiments". Ideally, a consistent solution of the FRG equation should also take into account mSTI's and mNI's controlling gauge and split symmetries in a coarse-grained way. In this sense, we expect that these symmetry identities will provide further constraints along the RG flow, eliminating the necessity of giving any physical meaning to such couplings arising from the symmetry-breaking terms induced by the regulator. A treatment involving the mSTI's and mNI's, however, goes beyond the scope of this paper.

\begin{center}
	\begin{table}[t]
		\begin{tabular}{|c|c|c|c|}
			\hline\hline
			& $\quad\qquad$ $\theta_1$ $\qquad\quad$  & $\quad\qquad$$\theta_2$$\quad\qquad$  & $\quad\qquad$$\theta_3$$\quad\qquad$ \\
			\hline
			%%%%%%%%%%%%%%%%%%%%%%%%%%%%%%%%%%%%%%%%%%%%%%%%%%%%%%%%%%%%%%%%%%%%%%%%%%%%%%%%%%%%%%%%%%%%%%%%%%
			$\quad$1-Loop \& $m_{k,i}^2\neq0$ -- Type I $\quad$ & $1.99 - 1.37 \,i$ & $1.99 + 1.37 \,i$ & $2.0$      \\
			1-Loop \& $m_{k,i}^2\neq0$ -- Type II         & $1.89 - 0.65 \,i$ & $1.89 + 0.65 \,i$ & $2.0$      \\
			%%%%%%%%%%%%%%%%%%%%%%%%%%%%%%%%%%%%%%%%%%%%%%%%%%%%%%%%%%%%%%%%%%%%%%%%%%%%%%%%%%%%%%%%%%%%%%%%%%
			Full System -- Type I      & $3.15 - 1.18 \,i$ & $3.15 + 1.18 \,i$ & $1.75$    \\
			Full System -- Type II     & $2.56 - 0.92 \,i$ & $2.56 + 0.92 \,i$ & $2.07$   \\
			\hline\hline
		\end{tabular}
	\caption{\footnotesize Critical exponents associated with the fixed point solutions reported in table \ref{Table_Fixed_Points}.}
	\label{Table_Crit_Exp}
	\end{table}
\end{center}

\begin{figure}[t]
	\begin{center}
		\includegraphics[height=6.2cm]{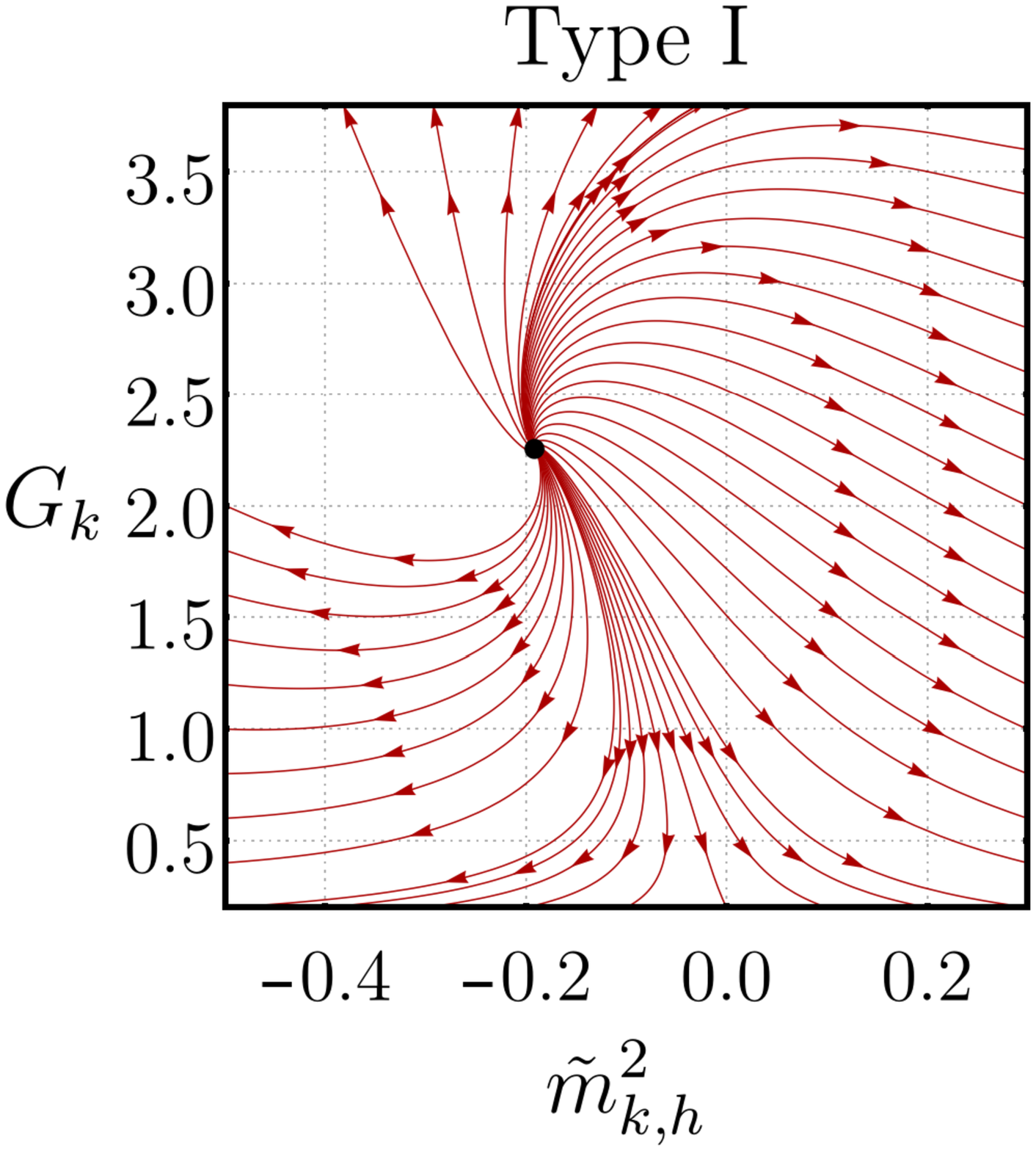} \quad\qquad
		\includegraphics[height=6.2cm]{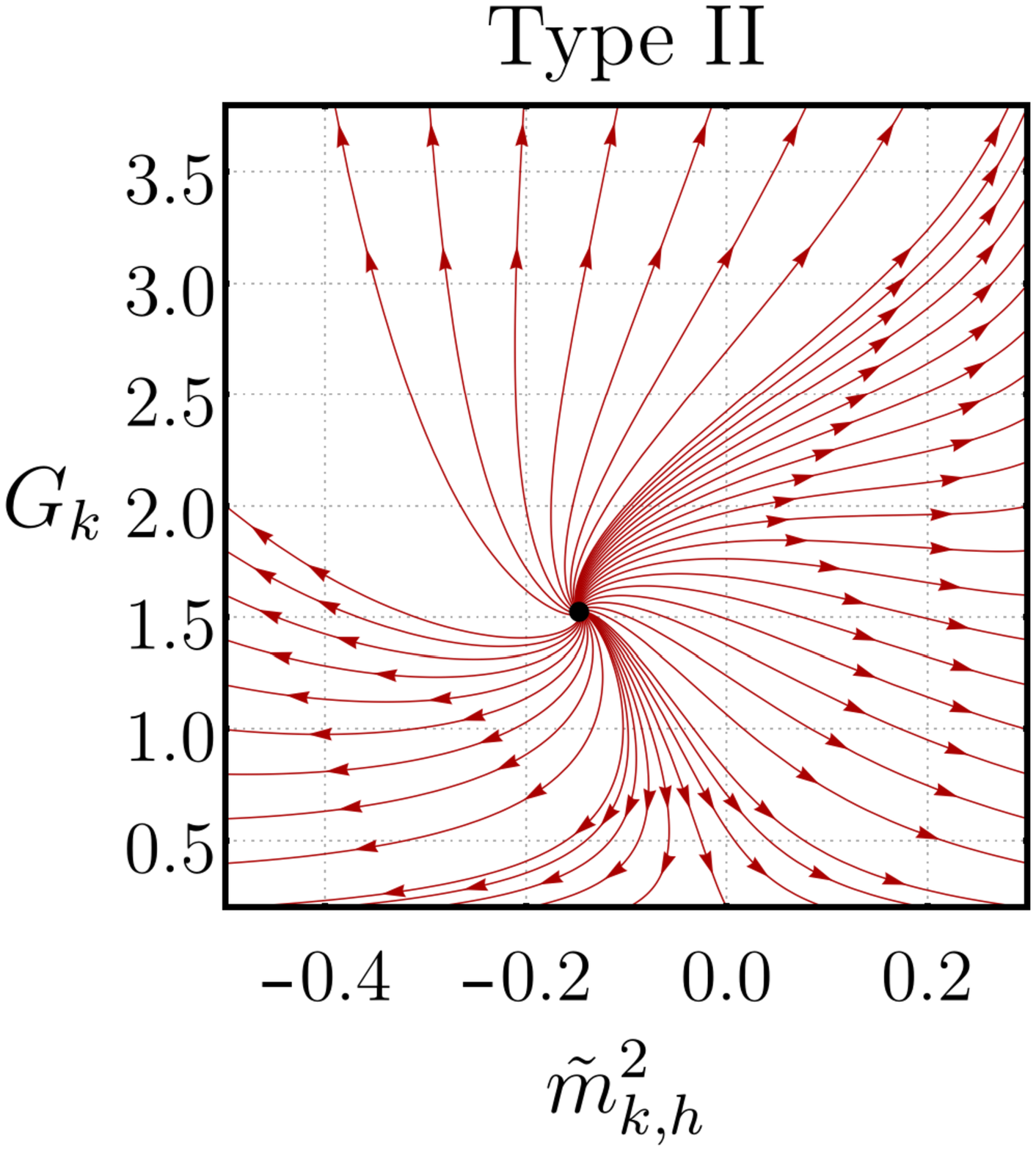} $\qquad$
		\caption{\footnotesize Flow diagram for unimodular quantum gravity in the single mass approximation. The RG trajectories correspond to numerical solutions of Eqs. \eqref{Beta_G_UG} and \eqref{mass_h}. As usual, the arrows point towards the infrared. In the single mass approximation, we found fixed point solutions $((\tilde{m}_h^2)^*,G^*)_{\textmd{Type I}} = (-0.19,2.25)$ and $((\tilde{m}_h^2)^*,G^*)_{\textmd{Type II}} = (-0.15,1.52)$, with corresponding critical exponents $\theta_{\pm}^\textmd{Type I} = 3.21 \pm 1.25 \,i$ and $\theta_{\pm}^\textmd{Type II} = 2.58 \pm 0.94 \,i$.}
		\label{Phase_Diagram}
	\end{center}
\end{figure}

Finally, in Fig. \ref{Phase_Diagram} we plot the RG flow diagram in unimodular quantum gravity. For simplicity we assume the single mass approximation $\tilde{m}_{k,\TT}^2 = \tilde{m}_{k,h}^2$ and $\tilde{m}_{k,\sigma}^2 = -2\tilde{m}_{k,h}^2$. In this case, the flow diagram corresponds to the integration of the system of equations \eqref{Beta_G_UG} and \eqref{mass_h} for various initial conditions in the $m_{k,h}^2 \times G_k$ plane. We should emphasize that the single mass approximation was adopted as way to avoid three dimensional plots and, therefore, it should be interpreted as an approximated slice of the truncated theory space defined by $G_k$, $\tilde{m}_{k,\TT}^2 = \tilde{m}_{k,h}^2$. Notably, the flow diagram represented in Fig. \ref{Phase_Diagram} exhibit remarkable similarities in comparison with the typical phase portrait in standard ASQG (see, e.g., \cite{Reuter:2001ag}), with the identification $\tilde{m}_{k,h}^2 = -2\tilde{\Lambda}_k$ ($\tilde{\Lambda}_k$ stands for the dimensionless cosmological constant). It is important to emphasize that the similarities between the flow diagrams for unimodular and standard ASQG do not necessarily imply the physical equivalence of these theories. In particular, the identification $\tilde{m}_{k,h}^2 = -2\tilde{\Lambda}_k$ does not take into account the different status of $\tilde{m}_{k,h}^2$ and $\tilde{\Lambda}_{k}$. As we have discussed in the previous paragraph, in unimodular ASQG, the symmetry breaking masses arise as an artifact induced by the FRG regulator and, therefore, we expect that the symmetry identities (mSTI's and mNI's) provide strong constraints such that the inclusion of symmetry breaking terms does not require additional initial conditions to be fixed by experiments. In the context of standard ASQG, the flow of the cosmological constant is not expected to be constrained by any symmetry identity, requiring initial conditions fixed by experimental observations.

\section{On the equivalence of Unimodular Gravity and Unimodular Gauge \label{Equiv_UGrav-UGauge} }

The unimodular gauge (sometimes also referred as physical gauge) has been used in the full \textit{Diff}-invariant version of ASQG as a convenient choice of gauge, \cite{Percacci:2015wwa,Ohta:2015fcu,Alkofer:2018fxj}. In the full-invariant case, we usually work with a gauge condition defined by the functional
\begin{align}
F_{\mu}[h\,;\bar{g}] = \bar{\nabla}^\nu h_{\mu\nu} - \frac{1+\beta}{4} \,\bar{\nabla}_\mu h^\tr \,,
\end{align}
where $\beta$ is a gauge parameter. Here we denote $h_{\mu\nu}$ as the full fluctuation field without the tracelessness condition. The unimodular gauge is characterized by a combination of the exponential parameterization ($g_{\mu\nu}=\bar{g}_{\mu\alpha} [\mathrm{e}^{\kappa\, h^{\cdot}_{\,\,\,\,\cdot}}]^{\alpha}_{\,\,\,\,\nu} $) with the limit $\beta \to -\infty$. This limit imposes a constant trace-mode $h^{\tr} = \textmd{const.}$, which, thanks to the exponential parameterization, imposes a kind of unimodular condition on the full metric $g_{\mu\nu}$. Note that this is \textit{a priori} very different from starting with a unimodular theory. In the present case, the tracelessness of the fluctuation arises as a gauge choice and, therefore, it is compensated by the inclusion of corresponding Faddeev-Popov ghosts. This is different from the unimodular setting, in principle, where no Faddeev-Popov ghosts are included in order to compensate for traceless fluctuations. In this sense, it is interesting to investigate whether the RG flow associated with the unimodular gauge provides equivalent results in comparison with the unimodular theory space explored in this paper. In this section, we present some arguments in favor of this equivalence. It is important to emphasize that all the statements in this section are valid at the level of the underlying FRG truncations. Moreover, we are considering pure gravity systems.

To perform practical calculations with the unimodular gauge we follow the same strategy to define a truncation as discussed in Sect. \ref{sect:setup}. In the present case we start from the ``seed'' truncation\footnote{For simplicity, we do not include the regulator induced masses. In this sense, we compare the RG flow in the unimodular gauge  with the case of unimodular gravity with $m_\TT^2=m_\sigma^2=0$. The discussion and conclusions presented here, however, can be extended to include the symmetry breaking masses in both settings.}
\begin{align}\label{seed_truncation_GR}
\hat{\Gamma}[h,c,\bar{c},b;\bar{g}] = 
\hat{\Gamma}_{\textmd{EH}}[g(h\,;\bar{g})] + \hat{\Gamma}_{\textmd{g.f.}}[h,c,\bar{c},b\,;\bar{g}] \, ,
\end{align}
where $\hat{\Gamma}_{\textmd{EH}}[g(h\,;\bar{g})] $ is the Einstein-Hilbert truncation
\begin{align}
\hat{\Gamma}_{\textmd{EH}}[g(h\,;\bar{g})] = 
\frac{1}{16\pi \GN} \int_x \sqrt{g}\, \left(2\Lambda - R(g(h\,;\bar{g}))\right) \,.
\end{align}
In contrast to unimodular gravity, we also include the cosmological constant term in the present case, since the metric determinant is not fixed \textit{a priori}. For the gauge-fixing sector, we consider the truncation
\begin{align}
\hat{\Gamma}_{\textmd{g.f.}}[h,c,\bar{c},b\,;\bar{g}] = 
\int_x \sqrt{\bar{g}}\, \, \bar{g}^{\mu\nu} \,b_{\mu} F_{\nu}[h\,;\bar{g}] -
\frac{\alpha\,}{2} \int_x \sqrt{\bar{g}}\,\bar{g}^{\mu\nu} \,b_{\mu} b_{\nu} + 
\int_x \sqrt{\bar{g}}\,\bar{c}_{\mu} \,{{\mathcal{M}}^{\mu}}_{\nu}[h\,;\bar{g}] \,c^{\nu}  \,.
\end{align}
Since the starting point corresponds to the full \textit{Diff}-invariant setting, the Faddeev-Popov ghost is not subject to the transversality condition as in unimodular gravity. Following the recipe discussed in Sect. \ref{sect:setup}, the dressed vertices are defined according to Eq. \eqref{dressed_vertices}, with slightly different dressing functions, namely
\begin{subequations}
	\begin{align}
	[\mathcal{Z}_{k,h}(p)^{1/2}]^{\mu\nu}_{\quad \alpha\beta} =\,\,
	&Z_{k,\TT}^{1/2} \, [\mathcal{P}_{\TT}(p)]^{\mu\nu}_{\quad \alpha\beta}
	+ \,\,Z_{k,\xi}^{1/2} \, [\mathcal{P}_{\xi}(p)]^{\mu\nu}_{\quad \alpha\beta} \\ \nn 
	+&\,Z_{k,\sigma}^{1/2} \, [\mathcal{P}_{\sigma}(p)]^{\mu\nu}_{\quad \alpha\beta} +
	Z_{k,\tr}^{1/2} \, [\mathcal{P}_{\tr}(p)]^{\mu\nu}_{\quad \alpha\beta} \,,
	\end{align}
	\begin{align}
	[\mathcal{Z}_{k,\bar{c}}(p)^{1/2}]^{\mu}_{\,\,\,\, \nu}  =
	[\mathcal{Z}_{k,c}(p)^{1/2}]^{\mu}_{\,\,\,\, \nu}  =
	Z_{k,c_\T}^{1/2} \, [\mathcal{P}_{\T}(p)]^{\mu}_{\,\,\,\, \nu} +
	Z_{k,c_\L}^{1/2} \, [\mathcal{P}_{\L}(p)]^{\mu}_{\,\,\,\, \nu} \,,
	\end{align}
	\begin{align}
	[\mathcal{Z}_{k,b}(p)^{1/2}]^{\mu}_{\,\,\, \nu} =
	Z_{k,b_\T}^{1/2} \, [\mathcal{P}_{\T}(p)]^{\mu}_{\,\,\, \nu}+
	Z_{k,b_\L}^{1/2} \, [\mathcal{P}_{\L}(p)]^{\mu}_{\,\,\, \nu} \,.
	\end{align}
\end{subequations}
The modification in the dressing functions accounts for the inclusion of the trace mode $h^\tr$ and also for the longitudinal sector of the Faddeev-Popov ghost and the Lautrup-Nakanishi field.

In this paper, we focus on the equivalence between unimodular gravity and unimodular gauge at the level of $n$-point connected correlation functions, with $n>1$, around flat background\footnote{More precisely, in terms of coarse-grained connected correlation functions obtained by taking functional derivatives of the scale-dependent Schwinger-like functional $W_k[J]$.}. In this sense, by looking at the FRG equations extracted from the $\beta$-dependent truncation defined above, the main result of this section can be summarized by the following equation
\begin{align}\label{Equivalence_Correlators}
\langle \varphi_{A_1}(p_1) \cdots \varphi_{A_n}(p_n) \rangle_k^{\textmd{conn.}} = 
\langle \varphi_{A_1}(p_1) \cdots \varphi_{A_n}(p_n) \rangle_k^{\textmd{conn.}}|_{\textmd{UG}} + \mathcal{O}(\beta^{-1}) \,,
\end{align}
with $n>1$, where $\langle \,\cdots \,\rangle_k^{\textmd{conn.}}|_{\textmd{UG}}$ denotes the correlation function evaluated in unimodular gravity (the absence of the subscript UG indicates that the correlation function is evaluated in the full \textit{Diff}-invariant version). Therefore, in the limit corresponding to the unimodular gauge ($\beta \to -\infty$) we verify that both settings lead to the same correlation functions. The crucial point to justify Eq.\eqref{Equivalence_Correlators} is the observation that, in the large-$|\beta|$ limit, the dressed propagators associated with the truncation defined in this section deviate from the propagators obtained in the unimodular gravity setting by $\mathcal{O}(\beta^{-1})$ contributions, namely
\begin{subequations}
	\begin{align}
	\textbf{G}_{k}^{\mu\nu\alpha\beta}(p) = 
	\textbf{G}_{k,hh}^{\mu\nu\alpha\beta}(p)|_\textmd{UG} + \mathcal{O}(\beta^{-1}) \,,
	\end{align}
	\begin{align}
	\textbf{G}_{k,c\bar{c}}^{\mu\nu}(p) = 
	\textbf{G}_{k,c\bar{c}}^{\mu\nu}(p)|_\textmd{UG} + \mathcal{O}(\beta^{-1}) \,,
	\end{align}
	\begin{align}
	\textbf{G}_{k,hb}^{\mu\nu\alpha}(p) = 
	\textbf{G}_{k,hb}^{\mu\nu\alpha}(p)|_\textmd{UG} + \mathcal{O}(\beta^{-1}) \,,
	\end{align}
	\begin{align}
	\textbf{G}_{k,bb}^{\mu\nu}(p) = 
	\textbf{G}_{k,bb}^{\mu\nu}(p)|_\textmd{UG} + \mathcal{O}(\beta^{-1}) \, .
	\end{align}
\end{subequations}
It is useful to consider the compact notation
\begin{align}\label{G_to_G_UG}
[\textbf{G}_{k}(p)]^A_{\,\,\,\,B} = 
[\textbf{G}_{k}^{\textmd{UG}}(p)]^A_{\,\,\,\,B} + \mathcal{O}(\beta^{-1})  \,.
\end{align}
As a consequence, in the unimodular gauge, the trace mode $h^\tr$ decouples and the longitudinal sector of the Faddeev-Popov ghost and the Lautrup-Nakanishi field as well. This result turns out to be sufficient to establish the equivalence between unimodular gravity and unimodular gauge. The basic idea is to express the connected correlation functions $\langle \varphi_{A_1}(x_1) \cdots \varphi_{A_n}(x_n) \rangle_k^{\textmd{conn.}}$ in terms of ``tree-level'' relations involving contractions of the dressed propagators and $n$-point vertices $\Gamma^{(n)}_{k,\,A_1 \cdots A_n}(\textbf{p})$ (with $n\geq3$). We take as an example the 3-point correlation function,
\begin{align}
\langle \varphi_{A_1}(p_1) \varphi_{A_2}(p_2) \varphi_{A_3}(p_3) \rangle_k^{\textmd{conn.}} =
[\textbf{G}_{k}(p_1)]^{B_1}_{\,\,\,\,A_1} \, [\textbf{G}_{k}(p_2)]^{B_2}_{\,\,\,\,A_2} \,
[\textbf{G}_{k}(p_3)]^{B_3}_{\,\,\,\,A_3} \,\Gamma^{(3)}_{k,\,B_1 B_2 B_3}(p_1,p_2,p_3) \,.
\end{align}
Using Eq. \eqref{G_to_G_UG} to express the dressed propagator, we find
\begin{align}
&\langle \varphi_{A_1}(p_1) \varphi_{A_2}(p_2) \varphi_{A_3}(p_3) \rangle_k^{\textmd{conn.}} =\,\, \nn \\ 
&\qquad\qquad\qquad =\,
[\textbf{G}_{k}^{\textmd{UG}}(p_1)]^{B_1}_{\,\,\,\,A_1} \, [\textbf{G}_{k}^{\textmd{UG}}(p_2)]^{B_2}_{\,\,\,\,A_2} \,
[\textbf{G}_{k}^{\textmd{UG}}(p_3)]^{B_3}_{\,\,\,\,A_3} \,
\Gamma^{(3)}_{k,\,B_1 B_2 B_3}(p_1,p_2,p_3) +
\mathcal{O}(\beta^{-1}) \,.
\end{align}
By looking at the structure of the dressed propagators in the case of unimodular gravity, we observe $\textbf{G}_{k}^{\textmd{UG}}(p)$ satisfies the following relation
\begin{align}\label{Proj_G_UG}
[\textbf{G}_{k}^{\textmd{UG}}(p)]^A_{\,\,\,\,B} = 
[\textbf{G}_{k}^{\textmd{UG}}(p)]^A_{\,\,\,\,C} \,\textbf{P}^{C}_{\,\,\,B}(p) = 
\textbf{P}^{A}_{\,\,\,C}(p) \, [\textbf{G}_{k}^{\textmd{UG}}(p)]^C_{\,\,\,\,B}   \,,
\end{align}
where $\textbf{P}^{A}_{\,\,\,B}(p)$ denotes to the traceless projector $\mathcal{P}_{1-\tr}$ (see Eq. \eqref{Traceless_Proj}) if contracted with indices associated with the fluctuation field $h_{\mu\nu}$ and stands for the transverse projector $\mathcal{P}_{\T}$ if contracted with indices associated with Faddeev-Popov or Lautrup-Nakanishi fields. Hence, the 3-point correlation function can be written as
\begin{align}
\langle \varphi_{A_1}(p_1) \varphi_{A_2}(p_2) \varphi_{A_3}(p_3) \rangle_k^{\textmd{conn.}} =\,\,&
[\textbf{G}_{k}^{\textmd{UG}}(p_1)]^{B_1}_{\,\,\,\,A_1} \, [\textbf{G}_{k}^{\textmd{UG}}(p_2)]^{B_2}_{\,\,\,\,A_2} \,
[\textbf{G}_{k}^{\textmd{UG}}(p_3)]^{B_3}_{\,\,\,\,A_3} \, \nn\\
&\!\!\!\!\times \textbf{P}^{C_1}_{\,\,\,B_1}(p_1) \textbf{P}^{C_2}_{\,\,\,B_2}(p_2)  
\textbf{P}^{C_3}_{\,\,\,B_3}(p_3) \,\Gamma^{(3)}_{k,\,C_1 C_2 C_3}(p_1,p_2,p_3) +
\mathcal{O}(\beta^{-1}) \,.
\end{align}
The action of $\textbf{P}^{A}_{\,\,\,B}(p)$ on $n$-point vertices essentially project out the trace mode $h^\tr$ and longitudinal components of $c^\mu$, $\bar{c}_\mu$ and $b_\mu$. Since the truncation defined in this section differs from the one discussed in Sect. \ref{sect:setup} by the presence of these modes, we can identify
\begin{align}
\textbf{P}^{B_1}_{\,\,\,A_1}(p_1) \,\cdots\, \textbf{P}^{B_n}_{\,\,\,A_n}(p_n)\,
\Gamma^{(n)}_{k,\,B_1 \cdots B_n}(\textbf{p}) =
\Gamma^{(n)}_{k,\,A_1 \cdots A_n}(\textbf{p})|_{\textmd{UG}} \,,
\end{align}
and, therefore
\begin{align}
&\langle \varphi_{A_1}(p_1) \varphi_{A_2}(p_2) \varphi_{A_3}(p_3) \rangle_k^{\textmd{conn.}} = \nn \\
&\qquad\qquad =\,\,
[\textbf{G}_{k}^{\textmd{UG}}(p_1)]^{B_1}_{\,\,\,\,A_1} \, [\textbf{G}_{k}^{\textmd{UG}}(p_2)]^{B_2}_{\,\,\,\,A_2} \,
[\textbf{G}_{k}^{\textmd{UG}}(p_3)]^{B_3}_{\,\,\,\,A_3} \, \Gamma^{(3)}_{k,\,B_1 B_2 B_3}(p_1,p_2,p_3)|_{\textmd{UG}} +
\mathcal{O}(\beta^{-1}) \, .
\end{align}
This result corresponds to the particular case with $n=3$ in Eq.\eqref{Equivalence_Correlators}. The same reasoning can be used to demonstrate Eq.\eqref{Equivalence_Correlators} for larger values of $n$. The case $n=2$ can be easily verified since the 2-point correlation functions corresponds to the dressed propagators itself. 

\begin{figure}[t]
	\begin{center}
		\includegraphics[height=2.0cm]{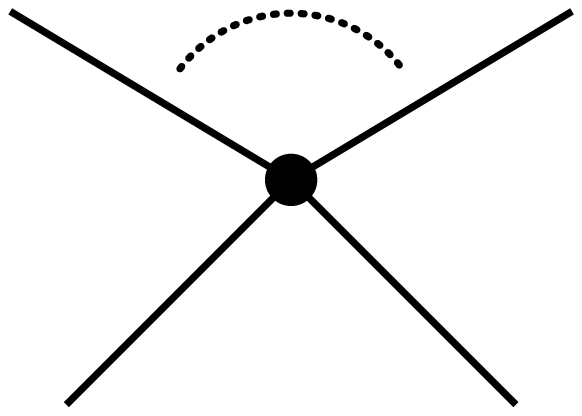} \quad\qquad\qquad\qquad
		\includegraphics[height=2.5cm]{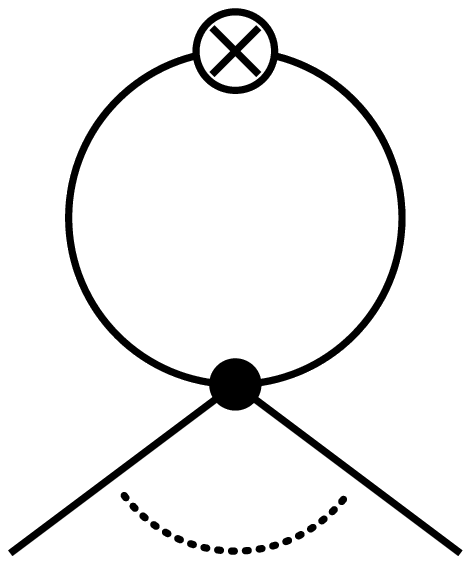} $\qquad\qquad\qquad$
		\put(-300,30){\Large{$\pt_t$}}
		\put(-180,30){\Large{$=$}}
		\put(-160,30){\Large{$-\frac{1}{2}$}}
		\put(-50,30){\Large{$+$}}
		\put(-35,30){\Large{$\,(\,\cdots\,)$}}
		%%%%%%%%%%%%%%%%%%%%%%%%%%%%%%%%%%%%%%%%%%%%%%%%%%%%%
		\put(-208,-10){$A_1$}
		\put(-275,-10){$A_2$}
		\put(-285,60){$A_3$}
		\put(-202,60){$A_n$}
		%%%%%%%%%%%%%%%%%%%%%%%%%%%%%%%%%%%%%%%%%%%%%%%%%%%%%
		\put(-130,-10){$A_1$}
		\put(-67,-10){$A_n$}
		\caption{\footnotesize Simplified diagrammatic representation of the flow equation for the $n$-point vertex $\Gamma_{k,\,A_1 \,\cdots \,A_n}^{(n)}(\textbf{p})$. Here we have used $(\,\cdots\,)$ to denote that there are other diagrams contributing to the flow equation.}
		\label{Diagramatic_Flow}
	\end{center}
\end{figure}

The argument presented above is sufficient to establish the equivalence of unimodular gravity and unimodular gauge at a fixed RG scale $k$. To complete the discussion, we still need to demonstrate that such an equivalence is preserved along the RG flow. For the dressed propagator, for example, it depends on the equivalence of the anomalous dimensions computed in both settings. In such a case, by means of computations performed within the truncation defined in this section we have explicitly verified that
\begin{subequations}
	\begin{align}
	\eta_\TT = \eta_\TT|_{\textmd{UG}} + \mathcal{O}(\beta^{-1}) \, ,
	\end{align}
	\begin{align}
	\eta_\sigma = \eta_\sigma|_{\textmd{UG}} + \mathcal{O}(\beta^{-1}) \,,
	\end{align}
	\begin{align}
	\eta_{c_\T} = \eta_c|_{\textmd{UG}} + \mathcal{O}(\beta^{-1}) \, ,
	\end{align}
\end{subequations}
and, therefore, in the we obtain the same result as if we start from unimodular gravity. To complete the discussion, we still need to show that
\begin{align}\label{Equivalence_Flow_Vertices}
\textbf{P}^{B_1}_{\,\,\,A_1}(p_1) \cdots \textbf{P}^{B_n}_{\,\,\,A_n}(p_n) \,
\pt_t \Gamma_{k,\,B_1 \,\cdots \,B_n}^{(n)}(\textbf{p}) = 
\pt_t \Gamma_{k,\,A_1 \,\cdots \,A_n}^{(n)}(\textbf{p})|_{\textmd{UG}} + \mathcal{O}(\beta^{-1}) \,.
\end{align}
In such a case, the basic idea is to use the flow equation for the $n$-point vertex, schematically represented as
\begin{align}\label{Flow_Vertices}
\pt_t \Gamma_k^{(n)} = 
-\frac{1}{2} \STr\Big( \textbf{G}_k \, \Gamma_k^{(n+2)} \, \textbf{G}_k \, \pt_t \textbf{R}_k \Big)
+ (\,\cdots\,) \, ,
\end{align}
where $(\,\cdots\,)$ denotes additional traces involving contractions of the dressed propagator $\textbf{G}_k $, the regulator insertion $\pt_t \textbf{R}_k$ and vertices $\Gamma_k^{(m)}$ (with $3\leq m \leq n+1$). Diagrammatically, Eq. \eqref{Flow_Vertices} is represented by Fig. \ref{Diagramatic_Flow}. Contracting the flow equation \eqref{Flow_Vertices} with $\textbf{P}^{B}_{\,\,\,A}(p)$, on its r.h.s. we obtain projected external lines, but keeping unprojected internal legs contracted with the dressed propagator $\textbf{G}_k$. Thanks to Eqs.\eqref{G_to_G_UG} and \eqref{Proj_G_UG}, in the limit $\beta \to -\infty$ the internal lines also become projected to the subspace defined by $\textbf{P}^{B}_{\,\,\,A}(p)$. With this observation, we can conclude that the projected vertices $\textbf{P}^{B_1}_{\,\,\,A_1}(p_1) \cdots \textbf{P}^{B_n}_{\,\,\,A_n}(p_n) \, \Gamma_{k,\,B_1 \,\cdots \,B_n}^{(n)}(\textbf{p})$ satisfy the same flow equations as the $n$-point vertices in unimodular gravity, justifying Eq. \eqref{Equivalence_Flow_Vertices} and completing our argument in favor of the equivalence of both settings at the level of $n$-point connected correlation functions, with $n>1$.

\section{Concluding Remarks}

Unimodular gravity is an equivalent description of the gravitional field at the classical level with respect to general relativity. The restriction of the symmetry group to volume-preserving diffeomorphisms does not reduce the dynamical content of the theory and all the gravitional phenomena described by general relativity can be accomodated in the unimodular setting. Conceptually, however, the cosmological constant arises as an integration constant in unimododular gravity and should be fixed by some initial condition. Quantum mechanically, summing over ``unimodular histories" can, in principle, produce differences with respect to the full \textit{Diff}-invariant theories framework.

In this work, we have analyzed the quantization of unimodular gravity within the asymptotic safety scenario for quantum gravity. In particular, we have shown that a UV fixed point exists - within the approximations implemented in this paper - when the RG equations are closed by computing the graviton and Faddeev-Popov ghosts anomalous dimensions independently by studying the flow of the two-point functions. Hence, our results give support to previous claims in the literature \cite{Eichhorn:2013xr,Eichhorn:2015bna,Benedetti:2015zsw} where a fixed point was found within a different approximation scheme. Besides that, by taking into account the considerations made in \cite{Ardon:2017atk,Percacci:2017fsy}, we have derived an explicit flow equation for unimodular gravity which takes into account a scalar ghost-like determinant which is generated by the functional measure. Such a new term does not introduce new vertices and, therefore, does not contribute to the flow of correlation functions.

Finally, we have performed a systematic comparison between unimodular gravity and the so-called unimodular gauge. It corresponds to a gauge-fixing condition for full \textit{Diff}-invariant theories where, in the exponential parameterization of the metric, the trace mode is decoupled. Our calculations, within the approximations that we used, reveal that unimodular gravity and full \textit{Diff}-invariant theories share the same connected $n$-point correlation functions (with $n>1$) and, in this sense, are equivalent. From the point of view of asymptotically safe quantum gravity, the cosmological constant appears as an essential coupling in full \textit{Diff}-invariant theories and, therefore, requires a fixed point to define a UV-complete theory. However, in the unimodular gauge, the cosmological constant decouples from all beta functions and just appears with the canonical scaling in its own beta function. Hence, the critical exponent associated to it is purely canonical and, therefore, corresponds to a relevant direction. Hence, there a free parameter associated to this direction in the theory space. In unimodular gravity, the cosmological constant is a free parameter that arises as an integration constant in the equations of motion. Consequently, although the cosmological constant does not appear in the ``unimodular theory space" it is still a free parameter that has to be fixed. It is worth mentioning that, in the unimodular flow, terms independent of curvature invariants are generated along the flow and if a running vacuum energy is added to the effective action, it will feature the same beta function as the cosmological constant in standard gravity. Possibly, this term can be properly taken into account by a suitable normalization of the flow, see, e.g., \cite{Lippoldt:2018wvi}. In summary, by looking at those theories as a collection of correlation functions, we are not able to distinguish them and the cosmological term seems to add just a free parameter in both cases.

This paper paves the way for the analysis of more sophisticated truncations of the effective average with the closure of the system with the independently-computed anomalous dimensions, see \cite{debrito2020}. Moreover, the interplay with matter seems to be a crucial point to be studied. Since, the coupling with matter does not contain the volume form, it is expected that matter-graviton vertices can show differences between unimodular theories and full \textit{Diff}-invariant ones, if any, see, e.g. \cite{deBrito:2019umw,Gonzalez-Martin:2017bvw,Gonzalez-Martin:2017fwz,Gonzalez-Martin:2018dmy,Herrero-Valea:2020xaq}.

\section*{Acknowledgments}

The authors thank A. Eichhorn, R. Percacci and A.F. Vieira for useful discussions on the topic. The authors also acknowledge M. Schiffer for various discussions concerning computational details and for providing expressions that allowed the cross-check of some of our results. GPB is grateful for the support by CAPES under the grant no.~88881.188349/2018-01 and CNPq no.~142049/2016-6 and thanks the ITP at Heidelberg University for hospitality. ADP acknowledges CNPq under the grant PQ-2 (309781/2019-1) and FAPERJ under the ``Jovem Cientista do Nosso Estado" program (E-26/202.800/2019).

\appendix

\section{Remarks on the Faddeev-Popov procedure in Unimodular Quantum Gravity \label{FlowEq_UG}}

The derivation of the flow equation from the path integral relies on a formal trick which is rather general. The introduction of a quadratic cutoff term supplemented with reasonable assumptions about the path integral measure lead to the exact flow equation with its one-loop structure \cite{Wetterich:1992yh} . In the case of gauge theories, one is confronted with the subtlety of introducing gauge-fixing conditions accompanied by Faddeev-Popov ghosts. Following the standard Faddeev-Popov procedure together with the standard prescription for the derivation of the flow equation, it is possible to derive the exact flow equation which takes into account the gauge-fixed version of the path integral.

In the special case of gravity, the derivation brings no further obstacles, with the exception that the background field method appears to be mandatory (but see \cite{Falls:2020tmj}). Hence, the derivation of the flow equation of diffeomorphism-invariant theories, the prescription is clear. Naively, there is no reason to expect that in the case of \textit{TDiff} being the gauge group the situation will change. However, the generators of \textit{TDiff} transformations are transverse contravariant vectors and transversality is a condition that depends on the choice of a metric. As pointed out in \cite{Ardon:2017atk,Percacci:2017fsy}, this entails some subtleties in the Faddeev-Popov procedure. In particular, the standard factorization of the gauge-group volume requires some non-trivial manipulations. In particular, there is a non-trivial contribution arising from the measure that will render a contribution to the flow equation, giving rise to the form \eqref{flowequnim} The (Euclidean) path integral of unimodular gravity can be written in terms of the traceless fluctuations $h_{\mu\nu}$ in the exponential decomposition, i.e.,
\begin{equation}
\mathcal{Z}_{\mathrm{UQG}} = \int \frac{\mathcal{D}h_{\mu\nu}}{V_{\mathrm{TDiff}}} \mathrm{e}^{-S_{\mathrm{UG}}(\bar{g};h)}\,,
\label{apa1}
\end{equation}
where $V_{\mathrm{TDiff}}$ corresponds to the volume of the \textit{TDiff} group and $S_{\mathrm{UG}}(\bar{g};h)$ corresponds to some classical unimodular action written in terms of the background metric $\bar{g}_{\mu\nu}$ and the fluctuation $h_{\mu\nu}$ - which arises due to the exponential decomposition. Following the standard Faddeev-Popov prescription, one inserts an identity in the path integral as
\begin{equation}
\mathcal{Z}_{\mathrm{UQG}} = \int \frac{\mathcal{D}h_{\mu\nu}}{V_{\mathrm{TDiff}}}\left(\int\mathcal{D}\epsilon^\mathrm{T}\Delta_{\mathrm{FP}}\,\delta (F^\mathrm{T})\right)\mathrm{e}^{-S_{\mathrm{UG}}(\bar{g};h)}\,,
\label{apa2}\,,
\end{equation}
with $\Delta_{\mathrm{FP}}$ standing for the Faddeev-Popov determinant. The delta functional implements the gauge-fixing condition $F^\mathrm{T}=0$ and the integral is performed over all transverse contravariant vectors $\epsilon^{\mathrm{T}}$. As pointed our in \cite{Ardon:2017atk,Percacci:2017fsy} , such an integral does not correspond to $V_{\mathrm{TDiff}}$. Actually, one can show that
\begin{equation}
V_{\mathrm{TDiff}} = \mathrm{Det}^{-1/2}(-\bar{\nabla}^2)\int\mathcal{D}\epsilon^{\mathrm{T}}\,.
\label{apa3}
\end{equation}
We emphasize that the ``extra" determinant which is necessary to define the volume depends on the background metric. In fact, this is a consequence which arises from the identity $\nabla_\mu \epsilon^\mu = \bar{\nabla}_\mu \epsilon^\mu$ which is just valid for unimodular metrics. Hence, by taking into account \eqref{apa3} we have, formally,
\begin{equation}
\mathcal{Z}_{\mathrm{UQG}} = \int {\mathcal{D}h_{\mu\nu}}{\mathcal{D}\bar{c}_{\alpha}}{\mathcal{D}c^{\beta}}\,\mathrm{Det}^{1/2}(-\bar{\nabla}^2)\,\mathrm{e}^{-S_{\mathrm{UG}}(\bar{g};h)-S_{\mathrm{gf}}(\bar{g};h,\bar{c},c)}\,,
\label{apa4}
\end{equation}
where $S_{\mathrm{gf}}$ denotes the gauge-fixing action together withe the Faddeev-Popov ghosts $\bar{c}_\alpha$ and $c^{\beta}$ term. Hence, besides the standard gauge-fixing term, one identifies the presence of the extra determinant in the path integral measure of \eqref{apa4}. Following the standard derivation of the flow equation, such an extra determinant can be regularized and gives rise to an extra contribution which can be effectively associated to a scalar ghost. Therefore, the flow equation in the case of unimodular gravity picks up a contribution from the path integral measure and leads to Eq.(\ref{flowequnim}). One important remark about such a modification is that, due to the unimodularity constraint, it just depends on the background and does not contain quantum fluctuations $h_{\mu\nu}$. Thence, in the so-called background approximation calculations, such a term will contribute and quantitatively affects the results regarding the fixed point structure. However, for the computation of the flow of $n$-point functions, this term automatically drops since functional derivatives with respect fluctuations, when acting on such a term, give a vanishing result.

\section{Explicit Results \label{Explicit}}

In this appendix we report the full expressions for the anomalous dimensions and symmetry breaking masses evaluated according to the projection rules defined in Sect. \ref{Flow2Pt}

\begin{align}\label{eta_TT_result}
\eta_\TT =& -\frac{5\,G_k\,
	( 468-120\, \tilde{m}_{k,\TT}^2 -696 \,\tilde{m}_{k,\TT}^4 + 
	(-43+73\,\tilde{m}_{k,\TT}^2 + 116 \,\tilde{m}_{k,\TT}^4) \,\eta_\TT ) }{2592\pi(1+\tilde{m}_{k,\TT}^2)^4}   \\
&+ \frac{G_k \, 
	( -441-816\, \tilde{m}_{k,\sigma}^2 -348\, \tilde{m}_{k,\sigma}^4 + 
	(73+131\,\tilde{m}_{k,\sigma}^2 + 58\, \tilde{m}_{k,\sigma}^4)\, \eta_\sigma )
}{648\pi (1+\tilde{m}_{k,\sigma}^2)^4} 
\nn \\
&-\frac{25\,G_k \,( - 16 - 8 \, \tilde{m}_{k,\TT}^2 -8\, \tilde{m}_{k,\sigma}^2 +
	(1+\, \tilde{m}_{k,\sigma}^2) \, \eta_\TT + (1 + \, \tilde{m}_{k,\TT}^2)\,\eta_\sigma )
}{576\pi \, (1+\tilde{m}_{k,\TT}^2)^2 (1+\tilde{m}_{k,\sigma}^2)^2}  + \frac{G_k\,(12-7\eta_c)}{96\pi} 
\nn \,,
\end{align}
%%%%%%%%%%%%%%%%%%%%%%%%%%%%%%%%%%%%%%%%%%%%%%%
%%%%%%%%%%%%%%%%%%%%%%%%%%%%%%%%%%%%%%%%%%%%%%%
\begin{align}\label{eta_sigma_result}
\eta_\sigma =& -\frac{5\,G_k \, 
	(-252-816\,\tilde{m}_{k,\TT}^2-132\,\tilde{m}_{k,\TT}^4 + 
	( 91+113 \,\tilde{m}_{k,\TT}^2 + 22 \,\tilde{m}_{k,\TT}^4 )\,\eta_\TT)
}{1296\pi\,(1+\tilde{m}_{k,\TT}^2)^4}  \\
&+\frac{G_k\, 
	(144+312\,\tilde{m}_{k,\sigma}^2-264\,\tilde{m}_{k,\sigma}^4 + 
	(-61-17\,\tilde{m}_{k,\sigma}^2+44\,\tilde{m}_{k,\sigma}^4)\,\eta_\sigma )
}{1296\pi\,(1+\tilde{m}_{k,\sigma}^2)^4} \nn \\
&+\frac{5\,G_k\, 
	(-16 - 8\,\tilde{m}_{k,\TT}^2 - 8\,\tilde{m}_{k,\sigma}^2 +
	(1 + \tilde{m}_{k,\sigma}^2)\,\eta_\TT + (1+\tilde{m}_{k,\TT}^2) \,\eta_\sigma)
}{144\pi \,(1+\tilde{m}_{k,\TT}^2)^2 (1+\tilde{m}_{k,\sigma}^2)^2}
- \frac{7\,G_k\,(4-\eta_c)}{24\pi} \,,\nn
\end{align}
%%%%%%%%%%%%%%%%%%%%%%%%%%%%%%%%%%%%%%%%%%%%%%%
%%%%%%%%%%%%%%%%%%%%%%%%%%%%%%%%%%%%%%%%%%%%%%%
\begin{align}\label{eta_Gh_result}
\eta_c =&\,\, \frac{5\,G_k \,(-24\,\tilde{m}_{k,\TT}^2 -5\,\eta_\TT 
	+ 3\,(1+\tilde{m}_{k,\TT}^2)\, \eta_c )}{648\pi \,(1+\tilde{m}_{k,\TT}^2)^2 } \\
&-\frac{G_k\,(-36-24\,\tilde{m}_{k,\sigma}^2 + \eta_\sigma +
	3\,(1+\tilde{m}_{k,\sigma}^2)\,\eta_c )}{81\pi \,(1+\tilde{m}_{k,\sigma}^2)^2} \,,\nn
\end{align}
%%%%%%%%%%%%%%%%%%%%%%%%%%%%%%%%%%%%%%%%%%%%%%%
%%%%%%%%%%%%%%%%%%%%%%%%%%%%%%%%%%%%%%%%%%%%%%%
%%%%%%%%%%%%%%%%%%%%%%%%%%%%%%%%%%%%%%%%%%%%%%%
%%%%%%%%%%%%%%%%%%%%%%%%%%%%%%%%%%%%%%%%%%%%%%%
\begin{align}\label{mass_TT}
\pt_t \tilde{m}_{k,\TT}^2 =& -(2-\eta_\TT)\,\tilde{m}_{k,\TT}^2 +
\frac{G_k\,(-620-1160\,\tilde{m}_{k,\TT}^2+(91+145\,\tilde{m}_{k,\TT}^2)\,\eta_\TT)
}{1296\pi \, (1+\tilde{m}_{k,\TT}^2)^3}  \\
&+\frac{G_k\,(100-440 \,\tilde{m}_{k,\sigma}^2 + (1+55\,\tilde{m}_{k,\sigma}^2) \,\eta_\sigma )}{6480\pi\, (1+\tilde{m}_{k,\sigma}^2)^3}
-\frac{G_k\,(110-7\eta_c)}{540\pi} \,,\nn
\end{align}
%%%%%%%%%%%%%%%%%%%%%%%%%%%%%%%%%%%%%%%%%%%%%%%
%%%%%%%%%%%%%%%%%%%%%%%%%%%%%%%%%%%%%%%%%%%%%%%
\begin{align}\label{mass_sigma}
\pt_t \tilde{m}_{k,\sigma}^2 =& -(2-\eta_\sigma)\,\tilde{m}_{k,\sigma}^2 -
\frac{G_k\,(-620-1160\,\tilde{m}_{k,\TT}^2+(91+145\,\tilde{m}_{k,\TT}^2)\,\eta_\TT)
}{648\pi \, (1+\tilde{m}_{k,\TT}^2)^3}  \\
&-\frac{G_k\,(100-440 \,\tilde{m}_{k,\sigma}^2 + (1+55\,\tilde{m}_{k,\sigma}^2) \,\eta_\sigma )}{3240\pi\, (1+\tilde{m}_{k,\sigma}^2)^3}
+\frac{G_k\,(110-7\eta_c)}{270\pi} \,. \nn 
\end{align}

For the sake of completeness, we also include the flow equation for the dimensionless mass parameter $\tilde{m}_h^2$ (single mass approximation).
\begin{align}\label{mass_h}
\pt_t \tilde{m}_{k,h}^2 =& -(2-\eta_\TT)\,\tilde{m}_{k,h}^2 +
\frac{G_k\,(-620-1160\,\tilde{m}_{k,h}^2+(91+145\,\tilde{m}_{k,h}^2)\,\eta_\TT)
}{1296\pi \, (1+\tilde{m}_{k,h}^2)^3}  \\
&+\frac{G_k\,(100+880 \,\tilde{m}_{k,h}^2 + (1-110\,\tilde{m}_{k,h}^2) \,\eta_\sigma )}{6480\pi\, (1-2\tilde{m}_{k,h}^2)^3}
-\frac{G_k\,(110-7\eta_c)}{540\pi} \,.\nn
\end{align}
The single mass approximation was used in the derivation of the flow diagram exhibit in Sect. \ref{RGFlowAndFP}.

\section{Projectors on flat background \label{Proj_Flat}}

The transverse and longitudinal projectors (on vector fields) are defined, around flat background, in the standard way
\begin{align}
\P_\T^{\mu\nu}(p) = \delta^{\mu\nu} - \frac{p^\mu p^\nu}{p^2} \,
\qquad \textmd{and} \qquad
\P_\textmd{L}^{\mu\nu}(p) = \frac{p^\mu p^\nu}{p^2} \,.
\end{align}
For rank-2 symmetric tensors, we define the projection operators
\begin{subequations}
	\begin{align}
	\P_\TT^{\mu\nu\alpha\beta}(p) = 
	\frac{1}{2}\left( \P_\T^{\mu\alpha}(p) \P_\T^{\nu\beta}(p) +
	\P_\T^{\mu\beta}(p) \P_\T^{\nu\alpha}(p) \right) -
	\frac{1}{3} \P_\T^{\mu\nu}(p) \P_\T^{\alpha\beta}(p) \,,
	\end{align}
	\begin{align}
	\P_\xi^{\mu\nu\alpha\beta}(p) = 
	\frac{1}{2}\left( \P_\T^{\mu\alpha}(p) \P_\textmd{L}^{\nu\beta}(p) +
	\P_\T^{\mu\beta}(p) \P_\textmd{L}^{\nu\alpha}(p) +
	\P_\T^{\nu\beta}(p) \P_\textmd{L}^{\mu\alpha}(p) +
	\P_\T^{\nu\alpha}(p) \P_\textmd{L}^{\mu\beta}(p)  \right) \,,
	\end{align}
	\begin{align}
	\P_\sigma^{\mu\nu\alpha\beta}(p) = \frac{1}{12}\P_\T^{\mu\nu}(p)\P_\T^{\alpha\beta}(p) 
	-\frac{1}{4} \P_\T^{\mu\nu}(p) \P_\textmd{L}^{\alpha\beta}(p)
	-\frac{1}{4} \P_\textmd{L}^{\mu\nu}(p) \P_\T^{\alpha\beta}(p)
	+\frac{3}{4} \P_\textmd{L}^{\mu\nu}(p) \P_\textmd{L}^{\alpha\beta}(p) \,,
	\end{align}
	\begin{align}
	\P_\tr^{\mu\nu\alpha\beta} = \frac{1}{4} \bar{g}^{\mu\nu} \bar{g}^{\alpha\beta}\,.
	\end{align}
\end{subequations}
These projectors select the different components of the usual York decomposition \cite{York:1973ia}. For the purpose of the discussion presented in Sect. \ref{Equiv_UGrav-UGauge}, it is also useful to define the traceless projector
\begin{align}\label{Traceless_Proj}
\P_{1-\tr}^{\mu\nu\alpha\beta} = 
\frac{1}{2} ( \bar{g}^{\mu\alpha} \bar{g}^{\nu\beta} + \bar{g}^{\mu\beta} \bar{g}^{\nu\alpha} ) 
- \frac{1}{4} \bar{g}^{\mu\nu} \bar{g}^{\alpha\beta}\,.
\end{align}

\bibliography{refs}
\end{document}